\documentclass[journal=jpcbfk,manuscript=article]{achemso}

\usepackage{achemso}
\usepackage{soul}
\usepackage{graphicx}
\usepackage{dcolumn}
\usepackage{bm}
\usepackage{booktabs}
\usepackage{textcomp}
\usepackage{booktabs}
\usepackage{hhline}
\usepackage{xcolor}
\usepackage[version=3]{mhchem}
\usepackage{amsmath}
\usepackage{braket}

\usepackage{subfig}
\usepackage{caption}

\title[]{Stress Effects on Vibrational Spectra of a Cubic Hybrid Perovskite: A Probe of Local Strain}

\author{Kuntal Talit}
\affiliation{Department of Physics, University of California, Merced, 5200 N. Lake Rd., Merced, CA 95343}

\author{David A. Strubbe}
\email{dstrubbe@ucmerced.edu}
\affiliation{Department of Physics, University of California, Merced, 5200 N. Lake Rd., Merced, CA 95343}
\SectionNumbersOn
\date{\today}

\begin{document}

\begin{abstract}\label{Abstract}
Inhomogeneous strain may develop in hybrid organic metal-halide perovskite thin films due to thermal expansion mismatch with a fabrication substrate, polycrystallinity or even light soaking. Measuring these spatially varying strains is difficult but of prime importance for understanding the effects on carrier mobility, non-radiative recombination, degradation and other optoelectronic properties. Local strain can be mapped using the shifts in vibrational frequencies using Raman or infrared microscopy.
We use density functional theory to investigate the effect of uniaxial strain on the vibrations of pseudo-cubic methylammonium lead iodide (CH$_3$NH$_3$PbI$_3$), and identify the vibrational modes most favorable for local strain mapping (86 cm$^{-1}$, 97 cm$^{-1}$, 1457 cm$^{-1}$, and 1537 cm$^{-1}$) and provide calibration curves. We explain the origin of the frequency changes with strain using dynamical matrix and mode eigenvector analysis and study strain-induced structural changes. We also calculate mode Gr\"uneisen parameters, giving information about anharmonicity and anisotropic negative thermal expansion as recently reported for other phases. Our results provide a basis for strain mapping in hybrid perovskites to further the understanding and control of strain, and improve stability and photovoltaic performance.

\end{abstract}
\section{Introduction}
Organic-inorganic hybrid perovskites are promising materials for next-generation solar cell applications.\cite{green} They have a direct bandgap, \citep{Frohna2018,Umari2014RelativisticGC} high absorption coefficient\citep{Shirayama2016a} long diffusion length,\citep{Stranks341,Xing344} and large carrier mobility\citep{Wehrenfennig2014} which make them favorable for PV applications. In the last 10 years the record photoconversion efficiency (PCE) has increased drastically from 3.8\% \citep{Kojima2009} to 25.2\%.\citep{NREL} Owing to other benefits like ease of bandgap engineering,\citep{Eperon861,Lin2014} flexibility for use in portable electronic devices,\citep{FlexiblePVSK} tunability of transparency to light for tandem cells,\citep{bailie2015semi} and suitability for mass production\citep{roldan2014flexible} with a cheaper cost, perovskites have been the object of great interest in the solar cell industry recently. There are also other non-PV applications of perovskites, \emph{e.g.} spin-optoelectronic applications,\citep{ping2018spin} luminescent solar concentrators,\citep{nikolaidou2016hybrid} and light-emitting diodes. \cite{hkim}\par{}
Despite these favorable properties, hybrid perovskites suffer from serious instability due to moisture,\citep{niu2015review} heat\citep{conings2015intrinsic} and light,\citep{Joshi2016} which hinder their commercialization as solar cells. Although different encapsulation techniques can help to eliminate the moisture issue,\citep{lee2015formamidinium,li2015outdoor} degradation due to unavoidable exposure to heat and light is still a challenge that needs to be addressed. Different studies have suggested that strain plays an important role in both degradation and photovoltaic performance of perovskites.\citep{ong2015structural,Nie2016,zhao2017strained,faghihnasiri2017dft,zhang2018strain,bechtel2018octahedral,tsai2018,jones2019lattice,Zhu2019,slotcavage2016light} This strain arises in thin films via substrate thermal expansion mismatch, epitaxial lattice mismatch,\citep{chen2020strain} phase transitions, grain structure, and composition inhomogeneities, creating both global and local strain.\citep{jones2019lattice} Strain within the lattice can affect the carrier dynamics,\citep{Zhu2019} defect concentration, non-radiative recombination,\citep{jones2019lattice} bandgap\citep{zhang2018strain} etc. and decrease the PCE of the device. Strain can also decrease the activation energy of ion migration within the material which accelerates its degradation.\citep{zhao2017strained} Slow photocurrent degradation caused by deep trap states has been attributed to polarons, consisting of strain and reorientation of the organic ion, which form under light soaking.\citep{Nie2016} On the other hand, light soaking can also create lattice expansion which reduces residual strain within the lattice and help to increase efficiency.\citep{tsai2018} High compressive strain (more than 1\% for cubic CH$_3$NH$_3$PbI$_3$) can also be responsible for structural phase changes.\citep{faghihnasiri2017dft,zhang2018strain} To understand all these strain-induced behaviors, we need to  understand the structural and vibrational changes due to strain.  

\par{}


In this work, we study as a benchmark system the pseudo-cubic phase of methylammonium lead iodide (MAPI, CH$_3$NH$_3$PbI$_3$),\citep{huang2017understanding,jena2019halide} one of the best studied hybrid perovskites. Depending on temperature, MAPI exists in three different phases: orthorhombic at low temperature ($T<$ 161.4 K), tetragonal at room temperature (161.4 K $< T <$ 330.4 K) and cubic (or pseudo-cubic)\cite{even2015pedestrian} at high temperature ($T>$ 330.4 K). Both orthorhombic and tetragonal structures are based on $\sqrt{2} \times \sqrt{2} \times 2$ supercells of the cubic structure with four times the number of atoms in the unit cell.

We have chosen to begin with a study of the high-temperature cubic phase due to its simpler structure. Further work will study the orthorhombic and tetragonal structures, which clearly have their own individual characteristics. The cubic phase can be considered as a reference to describe the structures of the other phases.\citep{even2015pedestrian} It is not only significant at high temperatures, but also it can be stabilized at lower temperature in quantum dots via ligands on the surface,\citep{sarang2017stabilization} use of other organic cations \citep{wu2018cations} such as formamidinium, or substitution of Br for I.\citep{sheng2015methylammonium} Generally cubic phases are found to have higher PCE than the other phases.\citep{zhang2015photovoltaic} Due to the strong resemblance between the cubic, tetragonal, and orthorhombic phases' vibrational properties, particularly in the mid-frequency range,\citep{Brivio} we believe this work also gives insight into those phases.

Vibrational properties of all three phases have been extensively studied.\citep{Luan,qiu2019room,ledinsky2015raman,nakada2019temperature,perez2018raman,Brivio,perez2015vibrational,glaser2015infrared,leguy2016dynamic} The infrared (IR) and Raman spectra of all three phases show three distinct regimes of the vibrational frequencies, due to the Pb-I cage and coupled cage/methylammonium (MA) ion modes at low frequency, and MA ion at medium and high frequency.\citep{perez2015vibrational} There are only a few experimental results on vibrational spectroscopy available for the cubic structure\cite{Luan,nakada2019temperature,qiu2019room,leguy2016dynamic}. Others have argued that there is no Raman activity of the low-frequency PbI cage modes for cubic MAPI\citep{Y.Zhang} or cubic MAPbBr$_3$,\citep{matsuishi2004optical} based on the symmetry argument given for CsPbCl$_3$ and SrTiO$_3$.\citep{even2015pedestrian} Observation of small but nonzero Raman activity in this regime was attributed to disorder.\citep{Y.Zhang} However, even in a perfectly ordered cubic MAPI crystal the fact that we have different atomic coordinates from CsPbCl$_3$ means that the PbI cage phonon modes (by hybridization with the MA ion modes) can have other irreducible representations that are Raman active, as indeed we demonstrate in this work. Measurement of low-frequency Raman has been attributed to light-induced degradation products,\citep{Y.Zhang} but our results suggest the validity of at least some measurements in this regime, demonstrated by the agreement in frequencies in Fig. \ref{fig:Raman_IR_comp}.

 The detailed behavior of vibrational modes under applied tensile and compressive strain for these materials has not been studied. A high-pressure study of MAPbBr$_3$\citep{matsuishi2004optical} showed only broad features in the cubic phase in the low-frequency region, and clear peaks occurring after phase transition to tetragonal. However the related macroscopic Gr\"uneisen parameter $\gamma$, and the related thermal expansion, have been studied. For the tetragonal structure, a negative thermal expansion coefficient was measured along the [001] direction ($c$-axis).\citep{Chunyu,heiderhoff2017thermal} For the cubic structure, the Gr\"uneisen parameter was calculated and measured in previous studies\citep{Chunyu,Brivio} but not along all three crystallographic axes. $\gamma$ (including the individual mode Grüneisen parameters, also known as phonon deformation potentials\citep{anastassakis1993effect}) are of fundamental interest as a probe of anharmonicity, which is quite strong in perovskites\cite{fabini2020underappreciated,zhu2019mixed} and relates to phonon-phonon scattering and thermal transport.

Using strain effects on vibrations to measure local strain can help us to understand the material degradation and its performance instability. To measure local strain within a material, we want to focus on a length scale $\sim \mu$m or below. Most of the experiments done so far to measure strain in perovskites have used grazing incidence X-ray diffraction (GIXRD), normal XRD\citep{rolston2018engineering,tsai2018,Zhu2019,zhao2017strained} or the substrate curvature method \citep{rolston2018engineering} which probe large areas. Scanning XRD has been used for more local mapping, but requires a synchrotron.\citep{jones2019lattice} Another standard non-destructive technique used to characterize perovskites and other semiconductor thin films is Raman spectroscopy.\citep{Pistor} Raman microscopy, in which small areas are probed, can be used to measure the stress distribution within a material. This is a well established technique for crystalline Si (c-Si) in the semiconductor industry,\citep{de1996micro} and is used for 2D materials.\citep{rao2019spectroscopic} It can even be used in hydrogenated amorphous silicon, a disordered material with very broad peaks, as we showed in a previous theory-experiment collaboration.\citep{Strubbe} This method has resolution on the $\mu$m scale, or even down to nanometers with tip-enhanced Raman spectroscopy.\citep{su2016nanoscale} Synchrotron-based IR microscopy has also been used for local mapping of strain and related structural changes.\citep{lyu2019phonon,smith2019infrared}

In this paper, we analyze vibrations under compressive and tensile strain for cubic CH$_3$NH$_3$PbI$_3$ and determine the calibration data needed to gauge local strain. The paper is organized as follows. In section \textrm{II} we detail our computational methods of structural optimization, calculation of normal mode frequencies and theoretical framework to understand frequency shifts in terms of perturbations of the dynamical matrix and mode eigenvectors. In section \textrm{III}, we discuss the behavior of phonon modes under uniaxial strain, understand the different behaviors in terms of structural changes and dynamical matrix analysis, find the best possible modes to probe local strain using Raman spectroscopy, and calculate the Gr\"uneisen parameter and compare to other theory and experimental data. In section \textrm{IV}, we conclude and present the four best modes for measuring strain using Raman or IR spectra.\par{}

\section{Methods}

\subsection{Computational Details}
For structural optimization and phonon mode calculations we have used density functional theory (DFT) and density functional perturbation theory (DFPT) \cite{Baroni.RevModPhys.73.515} as implemented in Quantum ESPRESSO (version 6.1)\cite{giannozzi2017advanced,Giannozzi,QE}. The Brillouin zone is sampled using a half-shifted 6$\times$6$\times$6 Monkhorst-Pack grid with an energy cutoff of 100 Ry for the wave-functions. Atomic positions are optimized until the total force per atom is smaller than 1 meV/${\rm \AA}$; for the initial variable-cell relaxation, a 0.5 kbar stress convergence threshold is used. The Local Density Approximation\citep{LDA_PhysRevB.23.5048} (LDA) with the Perdew-Wang (PW) parametrization\citep{perdew1992accurate} is used for the exchange-correlation potential for all the calculations. Scalar relativistic Optimized Norm-Conserving Vanderbilt (ONCV) pseudopotentials\citep{Hamann} are used which treat Pb 5$d$ orbitals as valence. All the pseudopotentials are taken from Pseud$\bar{o}$ D$\bar{o}$j$\bar{o}$\citep{van2018pseudodojo,pseudodojo} (NC SR ONCVPSP v0.4) with standard accuracy. We have not considered spin-orbit coupling as it does not have much effect on interatomic interactions near equilibrium.\citep{Brivio} \par{}
The initial structure is taken from the work of Brivio {\it et al.},\cite{Structure,Brivio} with the cation oriented close to the [100] direction, which was found to be slightly favored in molecular dynamics. It can be difficult to obtain an optimized exact cubic structure of MAPI without distortion to other phases.\citep{Frohna2018} Some experiments show the structure as cubic,\citep{PND} others pseudo-cubic.\citep{baikie2013synthesis,stoumpos2013semiconducting} Experimental work reports the fast reorientation of the CH$_3$NH$_3^+$ ion and also the rotation of CH$_3$ and NH$_3$ groups along the C-N axis, and these rotations and reorientation are extremely fast ($\sim$14 ps).\citep{leguy2015dynamics} These motions can make a cubic symmetry averaged over space and time, but obviously cannot be captured in a static DFT calculation, which instead represents a single local snapshot. A periodic calculation imposes an artificial long-range order, but our dynamical matrix calculations suggest that this does not have an important effect on the vibrational properties (Sec. 3.3), consistent with the conclusion of Leguy {\it et al.}\citep{leguy2016dynamic} that dynamical disorder broadens peaks in MAPI but does not shift them significantly. Local asymmetry in the structure along with spin-orbit coupling has been calculated to cause splitting and an indirect gap.\citep{Whalley2017,McKechnie2018} A theoretical study showed that changes in the \textit{c/a} ratio are coupled to the orientation of the CH$_3$NH$_3^+$ ion.\citep{ong2015structural} Our structural optimization makes the structure pseudo-cubic, in agreement with other calculations.\citep{ong2015structural,Brivio} As in standard DFT calculations, our lattice is at zero temperature, and the high temperature at which the cubic phase is observed does not enter into the calculations.\par{}
The equilibrium structural parameters are reported in Table \ref{tab:structureparam}. More detailed information, including results with different functionals, are reported in Table S1, along with bandgaps in Table S2 and the LDA bandstructure in Fig. S1 in agreement with previous work.\cite{Giorgi2013} Optimized crystallographic angles $\alpha$, $\beta$ and $\gamma$ are 90$^\circ$, 88.8$^\circ$ and 90$^\circ$ respectively. Pb-I-Pb angles are 164.65$^\circ$, 163.46$^\circ$ and 173.92$^\circ$ along $a$, $b$ and $c$ axes, respectively. The C-N bond-length is found to be 1.47 ${\rm \AA}$ with average C-H and N-H bond lengths of 1.1 ${\rm \AA}$ and 1.05 ${\rm \AA}$. Due to the pseudocubic lattice, off-centering of the Pb atom, cation orientation, and distortion of the Pb-I cage, the structure has no symmetry,\citep{even2015pedestrian} even when we checked the structure without the CH$_3$NH$_3^+$ ion or with the H atoms removed. Our final relaxed cation orientation has the C-N bond lying in the (010) plane, at an angle of 23.3$^\circ$ to the $a$-axis ([100] direction).\par{}
\begin{table*}[t]
\setlength{\tabcolsep}{6pt}
\renewcommand{\arraystretch}{1.5}
\caption{Optimized cell parameters for cubic CH$_3$NH$_3$PbI$_3$ from DFT energy minimization and comparison with previous studies using DFT/PBEsol,\citep{Brivio,Structure} powder neutron diffraction (PND)\citep{PND} and single crystal XRD\citep{stoumpos2013semiconducting} methods.}
\label{tab:structureparam}
\begin{tabular}{lcccccc}
\hhline{=======}
 & \textbf{a (${\rm \AA}$)} & \textbf{b (${\rm \AA}$)} & \textbf{c (${\rm\AA}$)}& \textbf{$\alpha$} (deg.)& \textbf{$\beta$} (deg.) & \textbf{$\gamma$} (deg.)\\
 \hline
DFT/LDA & 6.163& 6.115 & 6.267 & 90.00&88.80&90.00\\
DFT/PBE &6.499&6.410&6.532&90.00&88.66&90.00\\
DFT/PBEsol &6.291&6.248&6.378&90.00&88.64&90.00\\
DFT/PBEsol\citep{Brivio,Structure} & 6.289 & 6.229 & 6.374 &90.00&88.74&89.99 \\
PND (352 K)\citep{PND} & 6.317 & 6.317 & 6.317 &90.00&90.00&90.00 \\
XRD (400 K)\citep{stoumpos2013semiconducting} & 6.311 & 6.311 & 6.316 &90.00&90.00&90.00 \\
\hhline{=======}

\end{tabular}
\end{table*}

\begin{figure}[h]
  \includegraphics[width=\linewidth]{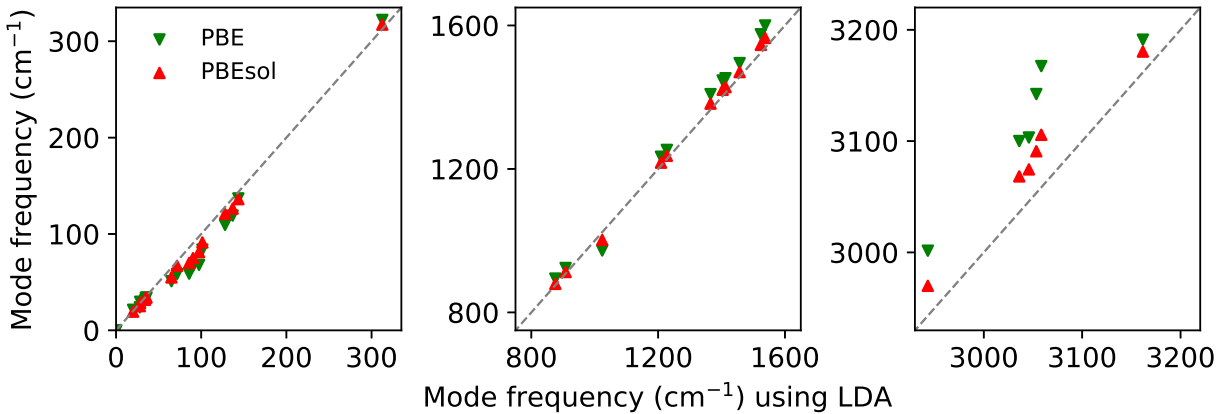}
  \caption{Comparison of $q=0$ phonon frequencies calculated with different functionals, showing good agreement across the low- and mid-frequency ranges, and some deviations at high frequencies.}
  \label{fig:PhononComp}
\end{figure}

Before applying strain to the structure, the convergence of the phonon frequencies with respect to the $k$-grid and phonon self-consistency threshold are checked, and all the calculations are done using phonon self-consistency threshold 10$^{-16}$ to avoid imaginary frequencies at the $\Gamma$ point. We compared different functionals (PBE,\citep{PBE_PhysRevLett.77.3865} PBEsol\citep{PBESol_PhysRevLett.100.136406} and LDA\citep{LDA_PhysRevB.23.5048}) and found only small changes in mode frequencies at $q=0$. For low-frequency phonons, the pattern is LDA$>$PBEsol$>$PBE whereas for medium (800--1600 cm$^{-1}$) and high (2900--3200 cm$^{-1}$) frequencies it is LDA$<$PBEsol$<$PBE with most deviation at high frequencies, as shown in Fig. \ref{fig:PhononComp}. For this comparison, we have performed the variable-cell relaxation using the same functional used to calculate the phonon frequencies.\par{}


For reference, we performed similar calculations for the well studied case of c-Si (Fig. S2) and the result follows the same low frequency pattern LDA$>$PBEsol$>$PBE that we have found in cubic MAPI. We apply uniaxial strain in the [100] direction and plot frequency changes; the three degenerate optical phonons are split into a doublet and singlet. Results are shown in Table S3 and Fig. S2. LDA in fact has the best agreement with experiment\citep{parker1967raman,anastassakis1990piezo} for both frequency and slope, in this case; PBEsol is similar, but PBE is more different.

Moving now to the effect of strain on cubic MAPI, we applied up to 1$\%$ uniaxial compressive (negative, in our convention) and tensile (positive) strain along the three crystallographic directions [100], [010] and [001] to the optimized zero-strain structure. Shear strains were not calculated as they generally have a smaller effect on the frequencies\citep{de1996micro}, and are less commonly found in thin films. We considered results up to $\pm$0.4$\%$ strain where the changes in frequency with the applied strain are typically at least 1 cm$^{-1}$, which is generally measurable in Raman spectroscopy.\citep{nakada2019temperature}
Strain more than $\pm$0.4$\%$ gives more signal-to-noise ratio, but more modes also become nonlinear. To make sure that this strain range is appropriate for studying linear responses, we have also computed the elastic constants by applying isotropic, tetragonal, and trigonal strain to the cubic MAPI lattice and thereby calculating elastic constants $C_{11}$, $C_{12}$ and $C_{44}$ following the procedure based on a quadratic fit of total energy\cite{giustino2014materials}. Within the range of $\pm$0.4$\%$, our result with LDA shows reasonable agreement with previously published results (Table \ref{tab:Elas-Const}). Note that we have separately calculated each diagonal element of the elastic constant tensor, and found some differences between them due to the pseudocubic lattice and lack of symmetry, whereas previous reports appeared to assume perfect cubic symmetry and calculated only $C_{11}$ and $C_{44}$.\citep{faghihnasiri2017dft,Feng2014}


\begin{table}[h]
\setlength{\tabcolsep}{5pt}
\renewcommand{\arraystretch}{1.5}
\caption{Calculated values of elastic constants in GPa for cubic CH$_3$NH$_3$PbI$_3$. }
\label{tab:Elas-Const}
\begin{tabular}{rrrrrrr}
\hhline{=======}
& \textbf{C$_{11}$} & \textbf{C$_{22}$} & \textbf{C$_{33}$} & \textbf{C$_{44}$} & \textbf{C$_{55}$} & \textbf{C$_{66}$}\\
\hline
\textbf{LDA (this work)} & 38.1 & 40.5 & 38.7 & 2.8 & 2.5 & 3.7\\
\textbf{PBE}\citep{Feng2014} & 27.1 & - & - & 9.2 & - & -\\
\textbf{PBEsol+vdW}\citep{faghihnasiri2017dft} & 35.4 & - & - & 6.1 & - & -\\
\textbf{PBEsol}\citep{faghihnasiri2017dft} & 30.9 & - & - & 3.2 & - & -\\
\hhline{=======}
\end{tabular}
\end{table}


Having confirmed reasonable results for LDA on phonon frequencies and elastic constants, we are using LDA for all our strain and phonon calculations as Quantum ESPRESSO can only provide Raman intensities for LDA.\citep{lazzeri2003first} The code's {\tt ASR=crystal} setting is used to enforce the acoustic sum rule (ASR) and make the acoustic modes exactly zero.\citep{Mounet} For each mode, we have calculated the uniaxial mode Gr{\"u}neisen parameter for [100], [010] and [001] crystallographic directions using the slope of the frequency versus strain graph, with the formula $\gamma_i=- \left( 1/\omega_i \right) d\omega_i / d\epsilon$, where $\epsilon$ is the applied strain and $\omega_i$ is the frequency of mode $i$\cite{Strubbe}. By taking the weighted average over all the modes, we have calculated the temperature-dependent Gr{\"u}neisen parameter $\gamma= \Sigma_i \gamma_i C_{v,i} / \Sigma_i C_{v,i}$. This is connected to the macroscopic property of the material by the relation $\gamma= \alpha K_T / C_v \rho$ where $\alpha$ is volume thermal expansion coefficient, $K_T$ is isothermal bulk modulus, $C_v$ is heat capacity at constant volume and $\rho$ is the density.\citep{Vocadlo1994} Raman intensities are calculated using the approach of Lazzeri {\it et al.}\citep{lazzeri2003first} and IR intensities computed from Born effective charge tensors which are calculated as the variation of macroscopic polarization with the atomic displacement using the modern theory of polarization\cite{Rabe,Baroni.RevModPhys.73.515}.

\subsection{Theoretical frame-work to understand the frequency change}
To obtain the normal mode frequencies, we solve the secular
equation\citep{Baroni.RevModPhys.73.515} given below:
\label{Section 2.2}
\begin{equation} \label{sec-eqn}
\sum_{J,\beta} D_{I\alpha, J\beta} u_{J\beta} = \omega^2 u_{I\alpha}
\end{equation}
where $I, J$ denote atoms within the unit cell; $\alpha, \beta$ represent $x, y$ and $z$ directions; $D_{I\alpha, J\beta}$ is the dynamical matrix of the system; and $u_{I\alpha}$ represents the mode eigenvector. The dynamical matrix can be expressed as
\begin{equation} \label{sec-eqn-massnormed}
D_{I\alpha,J\beta}=\frac{1}{\sqrt{M_I M_J}}\Big(\frac{\partial^2 E}{\partial R_{I\alpha}\partial R_{J\beta}}\Big)
\end{equation}
where $E$ is the total energy of the system, $R_I$ denotes the position vector of atom $I$, and $M_I$ denotes mass of atom $I$.
For non-strained condition, the normal-mode frequency $\omega$ will be represented as $\omega_{0}$. To treat the variation of $\omega$ with strain, we write $\omega^2$ as
\begin{equation} \label{seq}
\omega^2 =\sum\limits_{I \alpha, J \beta}{u^{*}_{I\alpha}D_{I\alpha, J\beta}u_{J\beta}}
\end{equation}

Strain can be treated as a perturbation to the vibrational properties of the material. The perturbation changes the dynamical matrix as well as the mode eigenvectors of the system, as given in equations (\ref{pdynmat}) and (\ref{eigV}) (up to quadratic order):

\begin{equation} \label{pdynmat} 
D^{\epsilon}_{I\alpha, J\beta}=D^0_{I\alpha, J\beta} + \sum\limits_{\l,m}\frac{\partial D_{I\alpha, J\beta}}{\partial \epsilon_{lm}}\epsilon_{lm}+\frac{1}{2}\sum\limits_{l,m,l',m'}\frac{\partial^2 D_{I\alpha, J\beta}}{\partial \epsilon_{lm}\partial \epsilon_{l'm'}}\epsilon_{lm}\epsilon_{l'm'} + \cdots
\end{equation}

\begin{equation}\label{eigV}
u_{J\beta}^\epsilon=u^0_{J\beta}+\sum\limits_{l,m}\frac{\partial u_{J\beta}}{\partial\epsilon_{lm}}\epsilon_{lm}+\frac{1}{2}\sum\limits_{l,m,l',m'}\frac{\partial^2 u_{J\beta}}{\partial \epsilon_{lm}\partial \epsilon_{l'm'}}\epsilon_{lm}\epsilon_{l'm'}+\cdots
\end{equation}

Since the mode eigenvectors are real in our case at $q=0$, we can use $u^*_{I\alpha}=u_{I\alpha}$. If we apply the perturbed dynamical matrix with perturbed eigenvector to equation (\ref{seq}), we can calculate the change in frequency as
\begin{equation}\label{changeFreq_eps2}
\begin{split}
\omega_{\epsilon}^2-\omega_{0}^2=\Large\sum\limits_{I\alpha, J\beta}\Large\sum\limits_{l,m} \Big\lbrace u^0_{I\alpha} \Big(\frac{\partial D_{I\alpha, J\beta}}{\partial\epsilon_{lm}}\Big) u^0_{J\beta} \Big \rbrace \epsilon_{lm} + \\
\Large\sum\limits_{I\alpha, J\beta}\Large\sum\limits_{l,m,l',m'} \Big\lbrace \frac{1}{2} u^0_{I\alpha}\Big( \frac{\partial^2 D_{I\alpha, J\beta}}{\partial\epsilon_{lm}\partial\epsilon_{l'm'}}\Big) u^0_{J\beta} +\Big(\frac{\partial u_{I\alpha}}{\partial\epsilon_{lm}}\Big) D^0_{I\alpha, J\beta} \Big(\frac{\partial u_{J\beta}}{\partial\epsilon_{l'm'}}\Big) +\\
2u^0_{I\alpha}\Big(\frac{\partial D_{I\alpha, J\beta}}{\partial\epsilon_{lm}}\Big)\Big(\frac{\partial u_{J\beta}}{\partial\epsilon_{l'm'}}\Big)+
u^0_{I\alpha} D^0_{I\alpha, J\beta}\Big(\frac{\partial^2 u_{I\alpha}}{\partial\epsilon_{lm}\partial\epsilon_{l'm'}}\Big) \Big\rbrace \epsilon_{lm} \epsilon_{l'm'} + \cdots
\end{split}
\end{equation}
The above equation can be used to find the change in frequency\citep{de1996micro} due to strain $\epsilon$ as
\begin{equation}\label{del_omega}
\Delta \omega \approx \frac{\omega_{\epsilon}^2-\omega_{0}^2}{2\omega_{0}}
\end{equation}
For silicon, a simple matrix equation for the Raman-active transverse optical (TO) modes has been theoretically derived\citep{anastassakis1993effect,ganesan1970lattice} and experimentally verified,\citep{ganesan1970lattice} but for the pseudo-cubic perovskite the secular equation is more complicated as it has five different types of atoms and has no symmetry or degeneracy. Instead, to analyze this situation, we have considered changes in eigenvectors with strain, as well as the changes in the dynamical matrices to clearly understand which atomic interactions contribute significantly to the changes in phonon frequencies, as detailed in section 3.3.\par{}


\section{Results and Discussion}

\subsection{Behavior of phonon modes under uniaxial strain}
Calculated normal modes at $q=0$ of cubic CH$_3$NH$_3$PbI$_3$ show three distinct frequency regions -- low (20-350 cm$^{-1}$), medium (850-1600 cm$^{-1}$) and high (2900-3200 cm$^{-1}$) -- as described for cubic as well as orthorhombic and tetragonal structures in previous studies.\citep{Brivio} Low-frequency modes are mainly due to the vibration of PbI$_6$ octahedra and some coupling between the PbI$_6$ octahedra and CH$_3$NH$_3^+$ ion, while the medium and high frequency modes involve the vibration of the CH$_3$NH$_3^+$ ion. Since pseudo-cubic MAPbI$_3$ does not have any symmetry,\citep{even2015pedestrian} it is not possible to assign spectral activity to any particular mode using group theory. All modes show both IR and Raman activity to some degree.\par{}
 
\begin{figure}[!ht]
   \centering
   \subfloat[][]{\includegraphics[width=.45\textwidth]{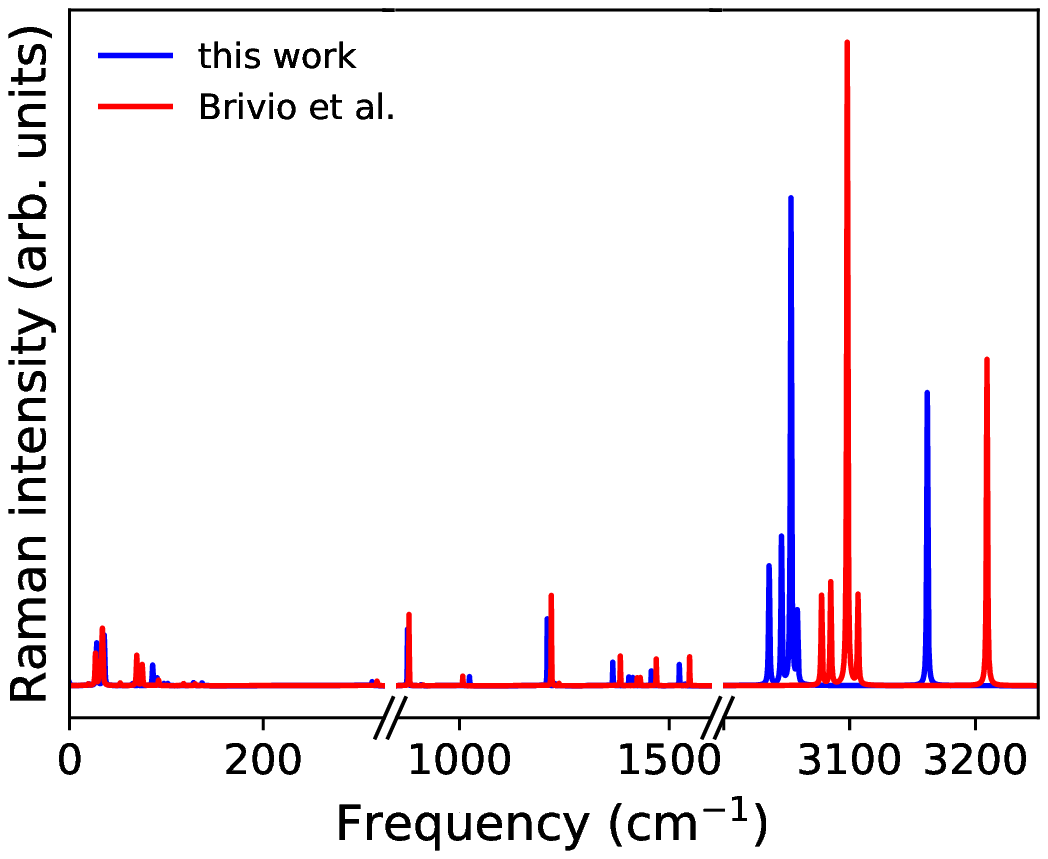}}\quad
   \subfloat[][]{\includegraphics[width=.45\textwidth]{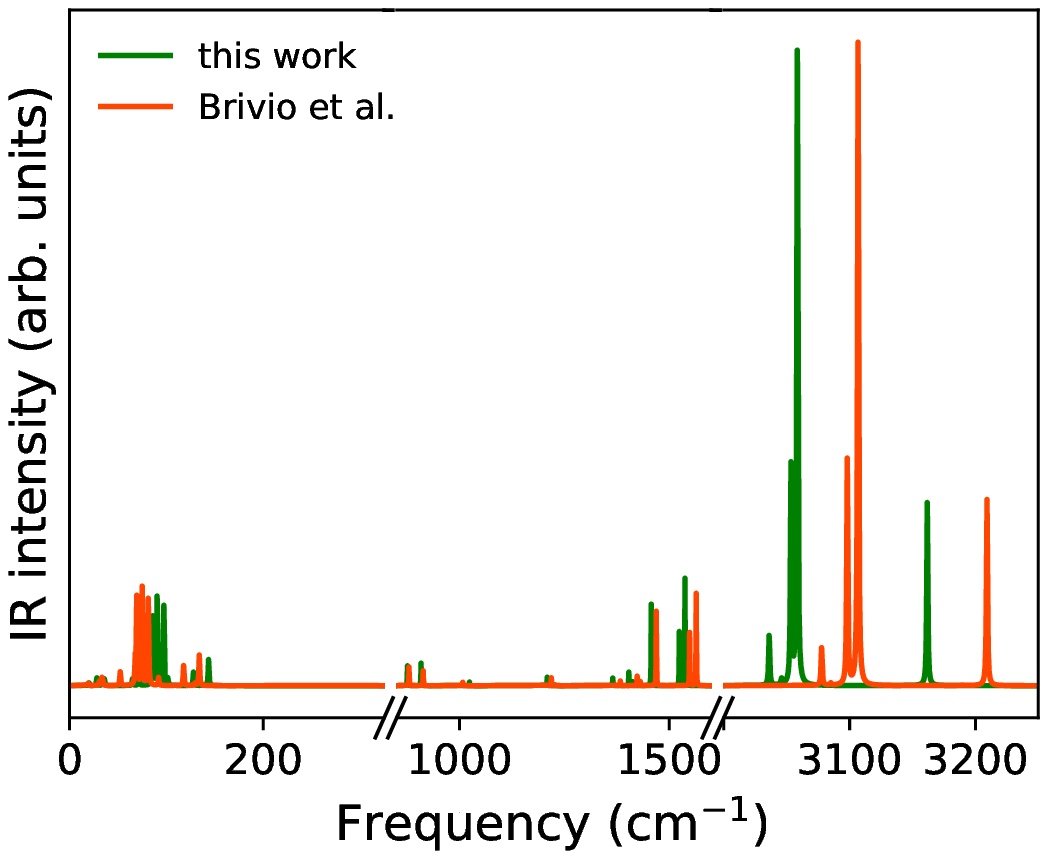}}\\
   \subfloat[][]{\includegraphics[width=.9\textwidth]{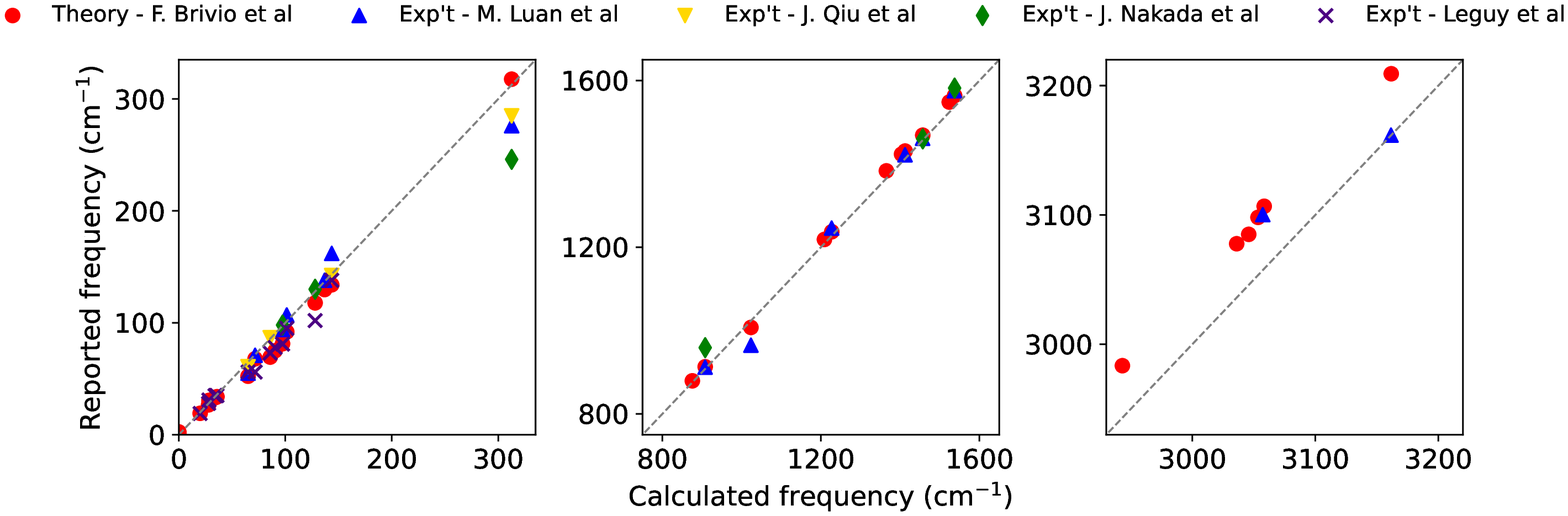}}
   \caption{Comparison of calculated Raman and infrared frequencies. (a) Comparison of Raman spectrum convolved with 1 cm$^{-1}$ Lorentzian broadening, with calculation of F. Brivio {\it et al.} \cite{Brivio} (b) Similar comparison for IR spectra. (c) Comparison of our calculated frequencies with published theoretical\citep{Brivio} and experimental results: IR in Luan \textit{et al.},\citep{Luan} Raman in Qiu \textit{et al.}\citep{qiu2019room} and Nakada \textit{et al.},\citep{nakada2019temperature} and THz in Leguy \textit{et al.}\citep{leguy2016dynamic}}
   \label{fig:Raman_IR_comp}
\end{figure}


\begin{figure}[h]
\centering
  \includegraphics[width=\linewidth]{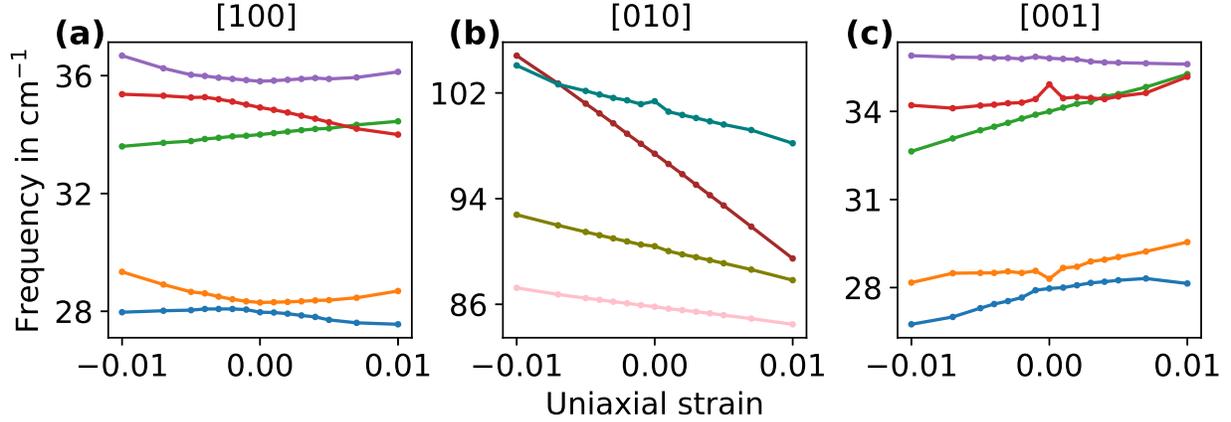}
  \caption{Frequencies of modes which approach or cross under strain, giving rise to parabolic or irregular behavior in frequency \textit{vs.} strain.}
  \label{fig:cross}
\end{figure}


 \begin{figure}[h]
\centering
  \includegraphics[width=\linewidth]{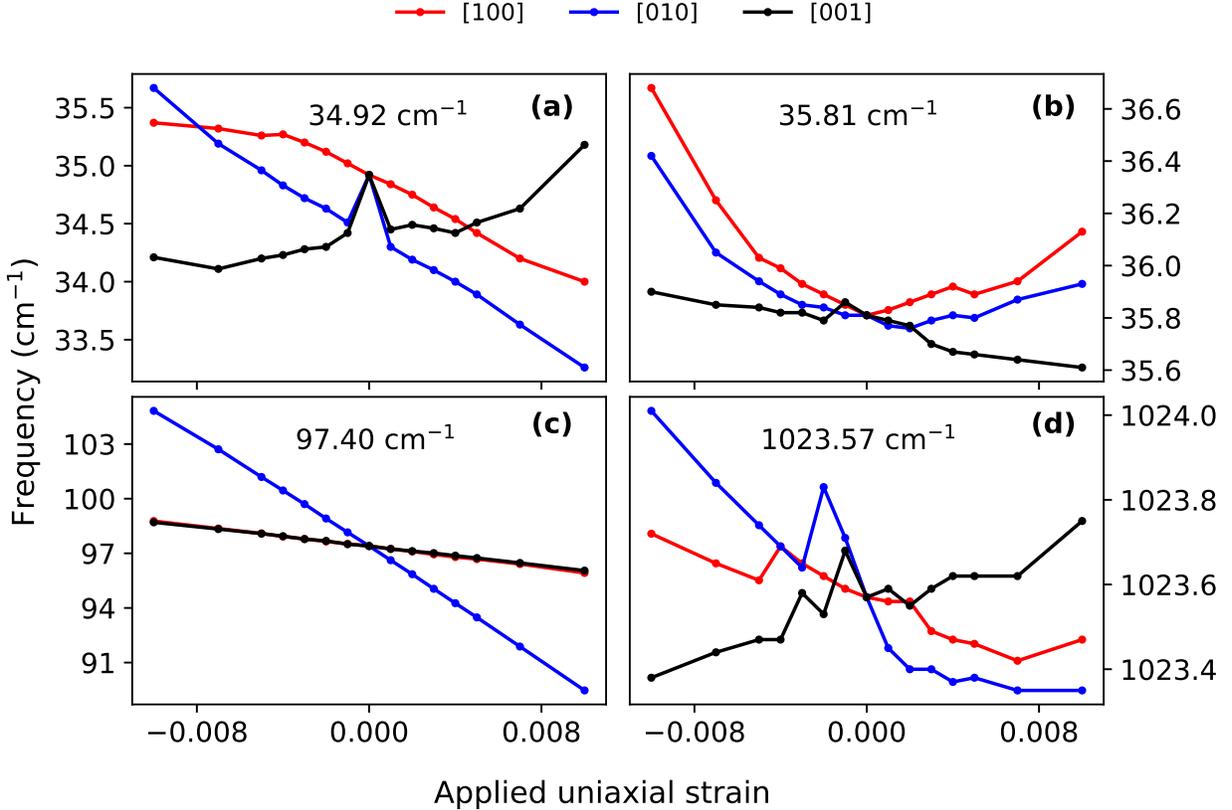}
  \caption{Typical patterns for frequency changes with applied compressive and tensile strain in three crystallographic directions [100], [010] and [001]. (a) Deep kink at zero strain for [010] and [001] but otherwise almost linear. (b) Parabolic pattern for [100] and [010] uniaxial strain. (c) Linear pattern. (d) Erratic pattern, neither linear nor parabolic. }
  \label{fig:Raman-shift}
\end{figure}

 \begin{figure}[h]
\centering
  \includegraphics[width=\linewidth]{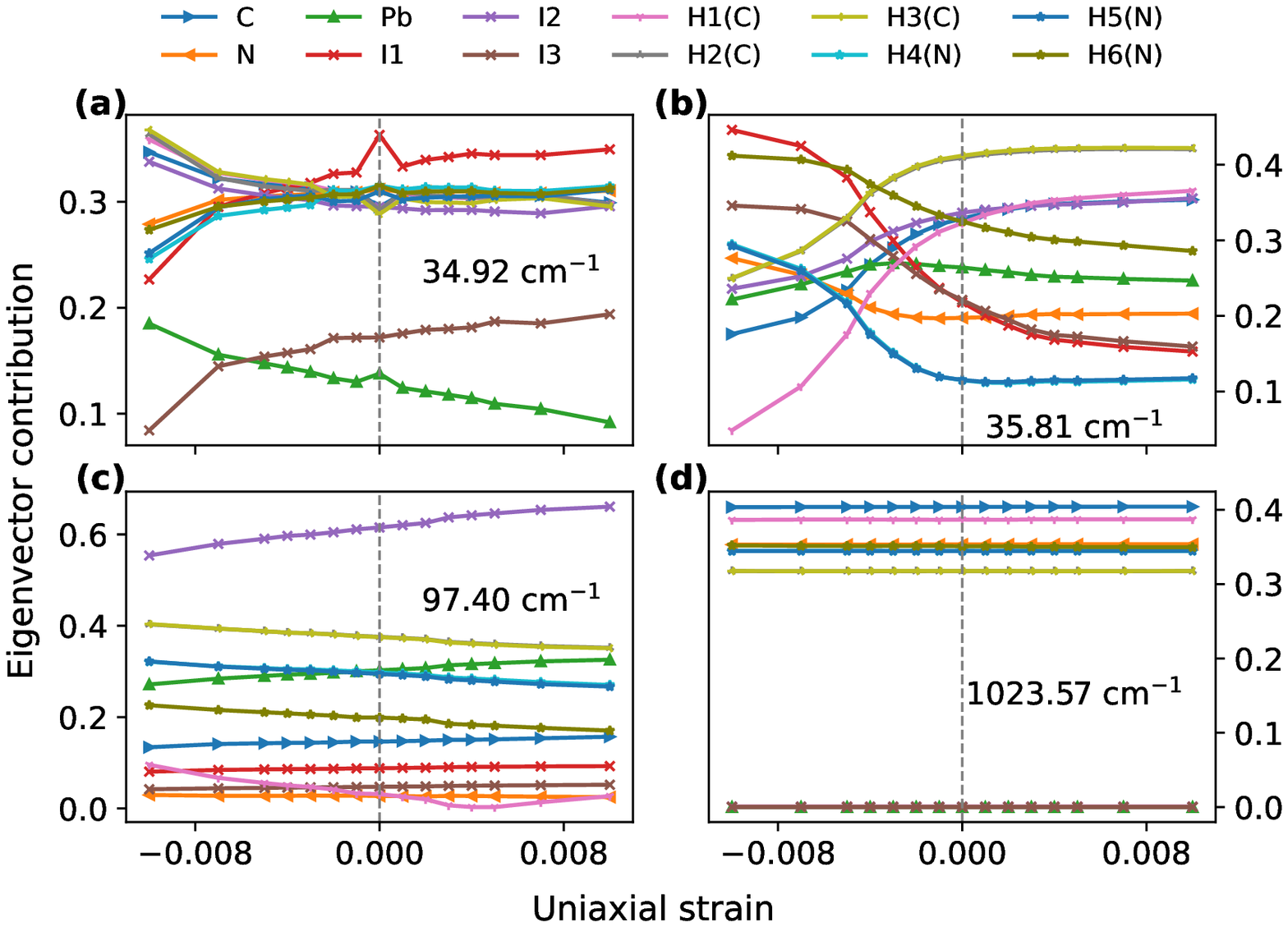}
  \caption{Contributions of each atom in the eigenvectors for each of the representative modes presented in Fig. \ref{fig:Raman-shift}. Uniaxial strains are in the (a) [010], (b) [100], (c) [001], and (d) [001] directions, showing connections between parabolic or other nonlinear frequency behavior in (a) and (b) with significant changes in mode character, while (c) has little character change and linear frequency behavior, and (d) anomalously has little character change but erratic frequency behavior which will be explained in \ref{subsection:structure}.
  \label{fig:EigenVec}}
\end{figure}

 Our {\it ab initio} IR and Raman results are shown in Fig. \ref{fig:Raman_IR_comp}. Calculated $\Gamma$-point phonon modes are convolved with 1 cm$^{-1}$ Lorentzian broadening and compared with the calculations of F. Brivio {\it et al.}.\citep{Brivio} The difference in results are primarily due to our use of LDA and their use of PBEsol, as expected from Fig. \ref{fig:PhononComp}. Calculated frequencies match well with published theoretical results,\citep{Brivio} and experimental measurements from THz at low frequencies \citep{leguy2016dynamic} and Raman\citep{nakada2019temperature,qiu2019room} and IR\citep{Luan} at medium to high frequencies.

 We have 12 atoms in our unit cell for cubic CH$_3$NH$_3$PbI$_3$, which gives 36 phonon modes, of which 3 have exactly zero frequency according to the ASR. For the other 33 modes we plotted frequency \textit{\textit{vs.}} applied strain (both compressive and tensile). Different patterns are noticed for different modes. Some of them are linear, some are parabolic, and some are neither (Fig. \ref{fig:Raman-shift}). For comparison, c-Si shows linear changes for small strain, and a splitting of degeneracy (Fig. S2). Given the much more complicated structure of the perovskite and lack of symmetry (hence no degeneracy), more complex behavior is observed. Low-frequency modes (27.97 cm$^{-1}$, 28.3 cm$^{-1}$, 34.92 cm$^{-1}$ and 35.81 cm$^{-1}$), which closely approach other modes under strain, show the parabolic pattern expected for an avoided crossing in phonons (or electronic bands), along with mixing and exchange of eigenvector character on either side of the crossing. The modes which approach or cross under strain are shown in Fig. \ref{fig:cross}. Using the eigenvectors as a guide, in a few cases we found crossing of modes under strain, and therefore relabeled the modes to maintain a continuous character. 
 Other modes with irregular behavior show even more mixing between modes, in a non-perturbative way, which we attribute to strong coupling to structural changes as discussed in Sec 3.2. Those modes where the frequency change is mostly linear can be categorized into four different categories: i) slopes of [100] and [010] strain are of same sign while slope of [001]  is opposite sign (\emph{e.g.} 143.6 cm$^{-1}$, 3053.38 cm$^{-1}$), ii) slopes of [010] and [001] strain are of same sign while slope of [100] is of opposite sign (\emph{e.g.} 908.09 cm$^{-1}$, 1403.54 cm$^{-1}$), iii) slopes of [100] and [001] strain are of same sign while slope of [010] is of opposite sign (\emph{e.g.} 20 cm$^{-1}$, 34 cm$^{-1}$) and iv) all slopes have same sign for [100], [010] and [001] strains (\emph{e.g.} 97.4 cm$^{-1}$, 312.59 cm$^{-1}$). The number of modes falling under different slope categories are given in Table S5 and the frequency change pattern for most of the modes is linear. There are some modes for which the slopes are almost the same for two different directions. For example, 97.4 cm$^{-1}$ has Pb-I-Pb bending perpendicular to both [100] and [001], giving rise to the same slopes for these strain directions. For 1365.3 cm$^{-1}$, symmetric umbrella type C-H bending has components in all directions and has comparable behavior for [100] and [010] strain directions. These behaviors indicate approximate symmetries of particular modes despite the lack of overall symmetry. In Fig. \ref{fig:Raman-shift}, we provide four representative modes' frequency \textit{vs.} strain patterns.


To understand these behaviors, we plotted eigenvectors (displacement patterns) for each mode, contributed by each atom in the unit cell as $\left|{u_I}\right|=\sqrt{(u_{Ix}^2+u_{Iy}^2+u_{Iz}^2)}$ ; those for the chosen four representative modes are given in Fig. \ref{fig:EigenVec}. We noticed that the change in frequency with strain is linear when the eigenvector does not change appreciably but the dynamical matrix does, as the effect of 2$^{nd}$ order change in dynamical matrix comes in only at 2$^{nd}$ order (Eqn. \ref{changeFreq_eps2}). Due to the lack of symmetry, the dynamical matrix elements always have a change with strain in this system. On the other hand, if the eigenvector does change, we can have some non-linear effect in the frequency change which can be understood from the second-order part of Eqn. \ref{changeFreq_eps2}. In most cases, the dynamical matrix change is large enough to produce effects beyond quadratic. For example,  Fig. \ref{fig:Raman-shift}(a) shows a kink at zero for the frequency for both [010] and [001] strain while for [100] strain it is almost linear (parabolic under compressive strain). We can see that the change in mode eigenvector for the corresponding mode shows a drastic change at zero for strain along [010] direction (Fig. \ref{fig:EigenVec}(a)) and the mode's character is a combination of translation of the CH$_3$NH$_3^+$ ion along [001] direction and Pb-I-Pb rocking mode. Fig. \ref{fig:Raman-shift}(b) shows parabolic frequency changes for [100] and [010] strain which can be understood from the changes of the eigenvector and close approach with another mode (Fig. \ref{fig:EigenVec}(b)). This mode is a combination of CH$_3$NH$_3^+$ ion libration in the [010] direction and a Pb-I-Pb rocking mode. In Fig. \ref{fig:Raman-shift}(c) we can see that the frequency change is linear and if we check the corresponding change in eigenvector (Fig. \ref{fig:EigenVec}(c)) we can see that there is very little change with strain which as per Eqn. \ref{changeFreq_eps2} will produce a linear pattern. The mode character for this mode is libration and spin (all the H atoms attached to C and N are rotating in the same direction) of CH$_3$NH$_3^+$ ion, and Pb-I-Pb bending. Finally, in Fig. \ref{fig:Raman-shift}(d) we found that the frequency change pattern is irregular despite very little change in atomic contributions to the eigenvector (Fig. \ref{fig:EigenVec}(d)). This is because the $x$, $y$ and $z$ components for C, N and H are changing irregularly with strain even though the sum of squares over the Cartesian directions is constant for each atom. This is a pure molecular mode with symmetric C-H and N-H bending with C-N stretch. We find that generally C-H and N-H vibrations are associated with nonlinear changes in the mode character under strain. We will see by structural and dynamical analysis that C-H, N-H vibrations are important for most of the frequency change under strain. Corresponding plots for all modes are given in Fig. S12. 
 
\subsection{Structural Changes under uniaxial strain}
\label{subsection:structure}

\begin{figure}[h]
\centering
  \includegraphics[width=\linewidth]{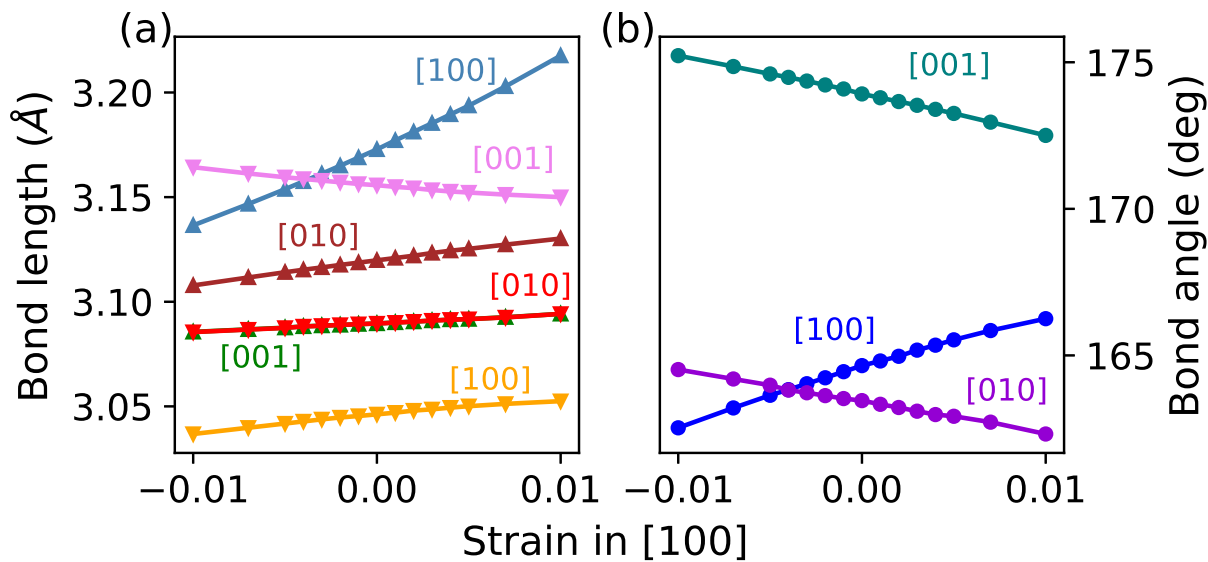}
  \caption{Change in Pb-I bond length and Pb-I-Pb bond angle in cubic CH$_3$NH$_3$PbI$_3$ for uniaxial [100] strain, showing fairly linear relationships, and a buckling of the Pb-I cage. These are the six Pb-I bonds, labeled by the crystal direction of the bond, and three Pb-I-Pb angles along [100], [010] and [001] directions respectively.}
  \label{fig:Bondlength-angle}
\end{figure}

While highly symmetric structures such as c-Si do not have internal parameters that can change with strain, in hybrid perovskites, the structure evolves under applied strain, which can play a role in vibrational changes. The significant changes are a sign of the anharmonicity of MAPI. Lengths of bonds parallel to strain are changed most, but those perpendicular to the direction of strain are also affected. For example, the Pb-I bond length along [100] changes most for strain along [100] but Pb-I bond lengths along other two perpendicular directions, such as along [010] and [001] are also changing as shown in Fig. \ref{fig:Bondlength-angle}(a). This change in bond lengths in the perpendicular direction to the strain is more prominent when the strain is acting perpendicular to the direction of the methylammonium ion which is close to [100] (Fig. S3). The Pb-I-Pb bond angles also change with applied strain: they decrease with compressive strain acting parallel to the bond angle and increase when it is acting perpendicular to the bond angle as can be seen in Fig. \ref{fig:Bondlength-angle}(b). Pb-I-Pb bond angle which is along [100] decrease, and those along [010] and [001] directions increase, for compressive strain along the [100] direction. Similar behavior is found in a theoretical study for tetragonal structure where the Pb-I-Pb bond angle which is parallel to the strain direction increases for tensile strain and decreases for compressive strain, but the Pb-I-Pb bond angle perpendicular to the strain direction increases for compressive strain and decreases for tensile strain.\citep{zhang2018strain} There are almost no changes ($<$0.007\%) in the C-N bond length but the N-H bond length (along [001]) decreases with compressive strain (Fig. S9). This reduction in N-H bond length supports the idea that compressive strain may be useful to stabilize the material.\citep{xue2020regulating,rappich2020light}

The CH$_3$NH$_3^+$ ion, which lies in the (010) or $xz$-plane also rotates with an increasing angle with respect to the [100] direction or $a$-axis under compressive strain and a decreasing angle under tensile strain (Fig. S11), with changes of up to $1.5^\circ$ over our strain range which is related to the change in $c/a$ ratio\citep{ong2015structural} as mentioned in Sec. 2.1. Rotation is largest for [010] strain. This may be due to the fact that the CH$_3$NH$_3^+$ ion lies in the (010) plane perpendicular to [010].
Rotation of CH$_3$NH$_3^+$ is also reported for tetragonal structure under compressive strain.\citep{zhang2018strain} The distances of C and N with their nearest Pb and I change with strains and the discontinuity in the I-N, I-C, N-Pb and C-Pb distances gives an indication why we see certain kinks or irregularities in the Raman shift vs strain graphs. For example, mode 128.01 cm$^{-1}$ shows kinks in the frequency change pattern (Fig. S12) where C-H and N-H are showing asymmetric bending modes with libration of the MA ion within the Pb-I cage, which is affecting the C-Pb and I-N distances most. Similarly, for 1023.57 cm$^{-1}$, we see that the irregularity is greater for [010] and [001] strain than [100]. We have further analyzed these changes in terms of the dynamical matrix in the next section. Full plots of structural parameters with each direction of strain are in Fig. S3-S11, and all atomic coordinates and phonon eigenvectors of relaxed strained structures are provided in the Supporting Information.
 \par{}

\begin{figure}[!ht]
   \centering
   \subfloat[][]{\includegraphics[width=.33\textwidth,height=2.5 in]{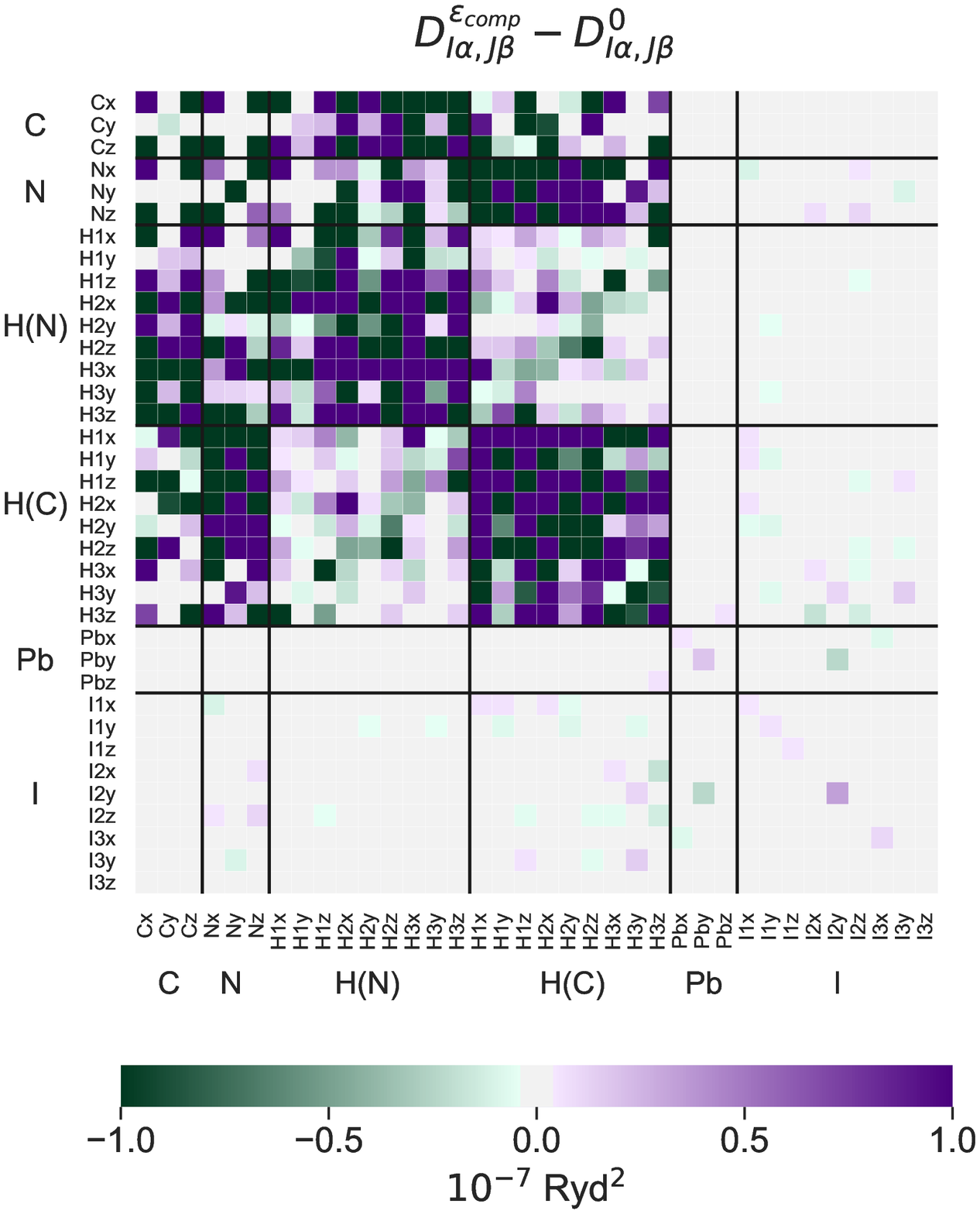}}
   \subfloat[][]{\includegraphics[width=.33\textwidth,height=2.5 in]{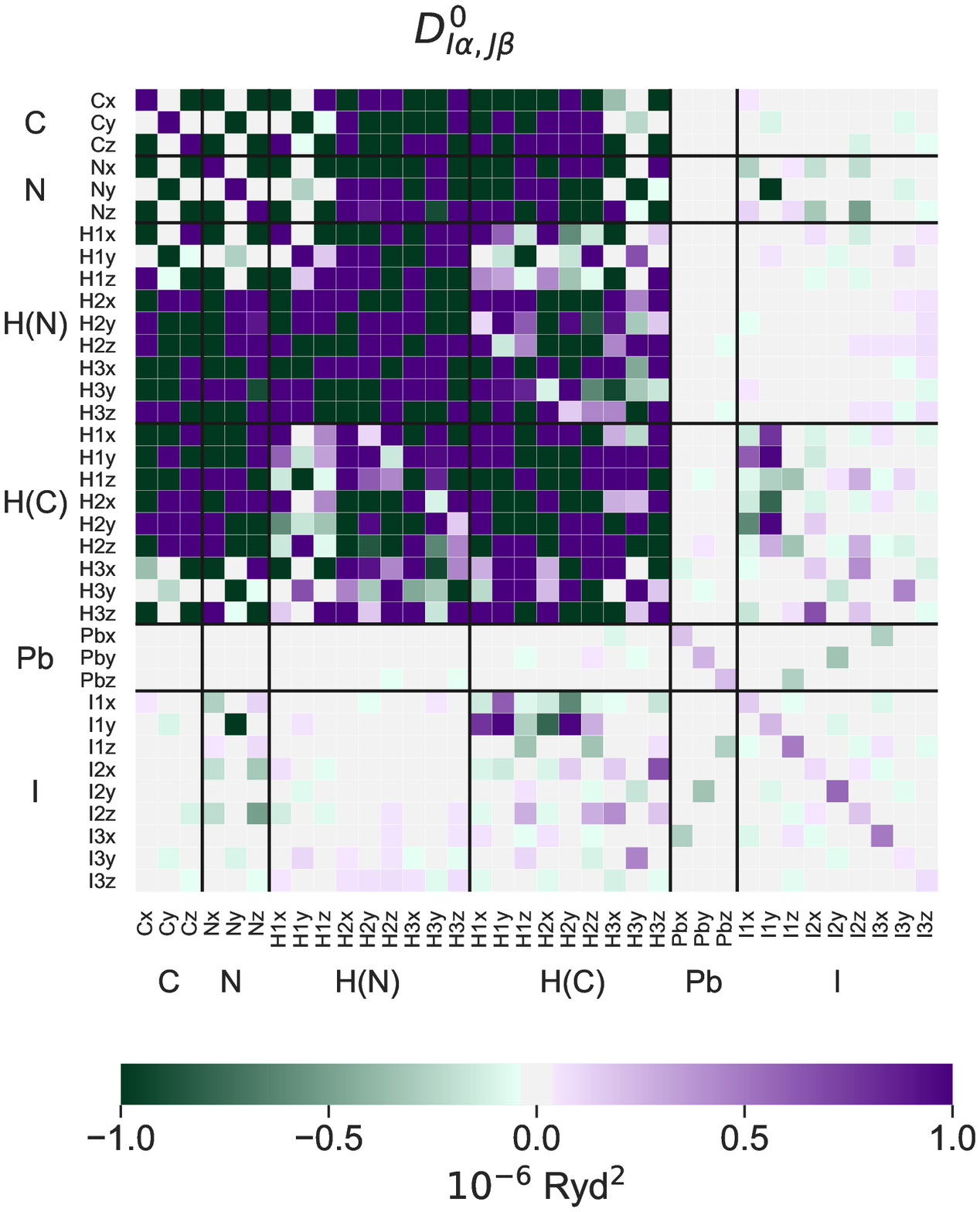}}
   \subfloat[][]{\includegraphics[width=.33\textwidth,height=2.5 in]{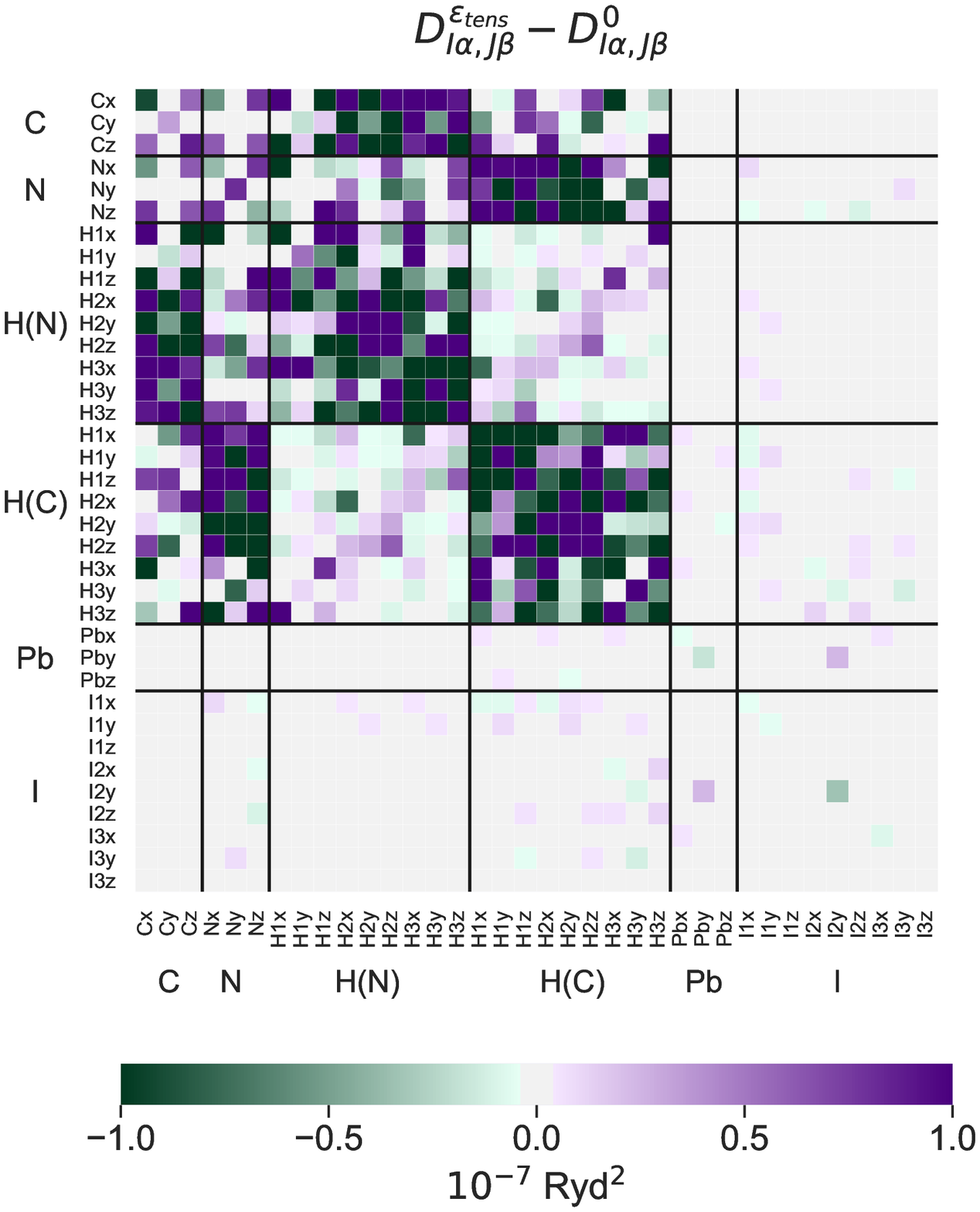}}
   \caption{Dynamical matrix for strain along [010] direction. (a) Change in dynamical matrix for compressive strain ($\epsilon = -0.004$). (b) Dynamical matrix at zero strain. (c) Change in dynamical matrix for tensile strain ($\epsilon = 0.004$). Symbols in both the axes represent atoms and their coordinates. For example, within H(N) block, H3z denotes the $z$ coordinate of the third H attached to N.}
   \label{fig:dynmat}
\end{figure}

\subsection{Change in dynamical matrix due to uniaxial strain.}
To understand the different behaviors of the phonon modes under strain, we have analysed the dynamical matrix of each strained structure according to Section 2.2. We have calculated the changes in dynamical matrices and plotted the results as heat-maps in Fig. \ref{fig:dynmat}(a) and \ref{fig:dynmat}(c) to understand which changes in interatomic interactions are most significant for the vibrational frequency changes under strain and how they relate to the different linear/parabolic/irregular pattern observed. Fig. \ref{fig:dynmat}(b) shows the dynamical matrix for the unstrained lattice. Note that these plots are symmetric about the diagonal and the scales are different for the strained and non-strained cases. The largest elements are in the C, N, H block due to lighter masses (see Eqn. \ref{sec-eqn-massnormed}). Diagonal elements are generally larger than off-diagonal elements, as can be seen at Fig. \ref{fig:dynmat}(b). It can be also seen that changes due to compressive strain are more than those of tensile strain, which can be understood based on the typical curve of energy \textit{vs.} bond length, $e.g.$ the Morse potential. 
Since Pb and I are heavier than C, N and H, Pb-I interactions will be significant most for low frequency modes. It can be noticed that there are very small interactions between H atoms and their nearest Pb or I atoms (Fig. \ref{fig:dynmat}(b)) which gives an indication that van der Waals interactions are of minor importance for vibrations of the cubic structure which is also supported by a previous study.\citep{kang2017preferential} We have calculated also the dynamical matrix for a $2 \times 2 \times 2$ supercell, and found that the matrix elements are significantly smaller when the two atoms are in different primitive cells. This indicates that the artificial long-range order in our periodic structure does not make much difference in the vibrational properties compared to the true dynamical disorder. \par{}
From the change in dynamical matrix in Fig. \ref{fig:dynmat}(a) and Fig. \ref{fig:dynmat}(c), we can see that the component of Pb-I interactions changes most when it is parallel to the direction of strain. For example, the Pb$_y$-I$_y$ interaction changes most for [010] strain whereas Pb$_x$-I$_x$ and Pb$_z$-I$_z$ interactions change most for [100] and [001] strains respectively (Fig. S14-S15). We have already seen in the structural changes that Pb-I bond length is affected most when the stretches are along the direction of polarization of the bond. It is clear from Fig. \ref{fig:dynmat}(a) and Fig. \ref{fig:dynmat}(c) that Pb-I modes are significantly affected due to strain. It can also be seen that I-N coupling (lower left and upper right) is important and greater than I-C coupling which is due to the electrostatic interaction between the MA ion and the cage. These interactions play a significant role in the frequency shifts and are important for the medium and high frequency regions. \par{}
Given the success of analyzing vibration separately\citep{perez2015vibrational} (lattice and ion), one might think that the lattice strain will not affect the MA ion, but this turns out not to be true. Among the hydrogens attached to C, or the hydrogens attached to N, there are significant changes with strain in the interactions, which explain the behaviors of some of the high frequency modes. Modes that have a large component of C-H, N-H stretch have higher slopes in the high frequency region, whereas in the low frequency region large Pb-I stretch or bend with C-H, N-H stretch is associated with large slope. C-H, N-H asymmetric stretch or CH$_3$NH$_3$ spin and torsion are found to have high slopes.
It is found that slope values are  high in the low and high frequency region and small in the medium frequency region. It is also found that the slope values are typically higher when the strain is perpendicular to the direction of the MA ion (along [010] and [001]) and lower when its along the direction of the MA ion \textit{i.e.} along [100]. We attribute this asymmetry to the orientation of the C-H bonds: two of the C-H and N-H bonds lie closer to [010] than to [100], giving greater effects from [010]; one C-H and one N-H bond lie parallel to [001], and these bond lengths are affected most by [010], and also show the greatest effects of any C-H or N-H bonds in any strain direction (Fig. S9, S10). Therefore the strain effects on the MA ion, and its interactions with the cage, are greatest when we apply strain along [010] and [001]. 

IR and Raman intensities are much higher in the high frequency region than in the low or mid-frequency regions, and are mainly contributed by C-H and N-H stretches. From our full phonon analysis, shown in Table S4, we found that low-frequency modes are mainly due to Pb-I-Pb vibration with translation or libration mode of the MA ion. The mid-frequency modes mainly consist of C-H and N-H asymmetric bending with some spin or twist. The high-frequency modes are mainly symmetric and asymmetric stretch of C-H and N-H bonds of the methylammonium ion. For all the modes, IR intensity vectors are lying in the plane, either in $xz$ (100) or in $xy$ (001) plane. Modes which have high slopes in all three directions -- among them, high frequency modes (3053.38 cm$^{-1}$, 3058.42 cm$^{-1}$, 3161.71 cm$^{-1}$) -- are both highly IR- and Raman-active while low frequency modes (27.97 cm$^{-1}$, 65.02 cm$^{-1}$, 85.82 cm$^{-1}$) are less IR- and Raman-active. Modes involving libration and translation of the CH$_3$NH$_3^+$ ion and Pb-I-Pb bending have high slopes along [100] direction while modes having high slope in [010] show C-H and N-H stretch in CH$_3$NH$_3^+$ ion. 

\subsection{Best modes to probe local strain in cubic CH$_3$NH$_3$PbI$_3$}


\begin{figure}[h]
\centering
  \includegraphics[width=\linewidth]{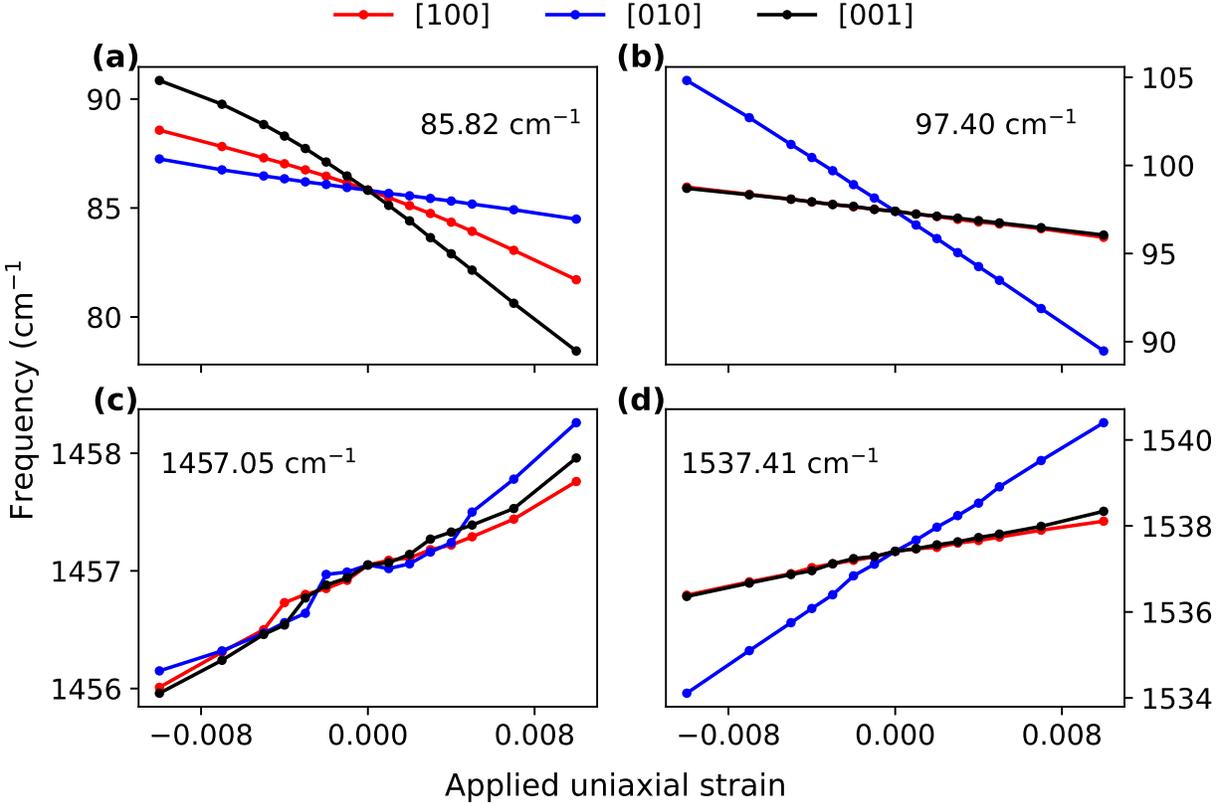}
  \caption{Frequency \textit{vs.} strain for the best modes for IR/Raman microscopy to probe local strain, showing linear changes for robust calibration.}
  \label{fig:bestRamanmode_freq}
\end{figure}

\begin{figure}[h]
\centering
  \includegraphics[width=\linewidth]{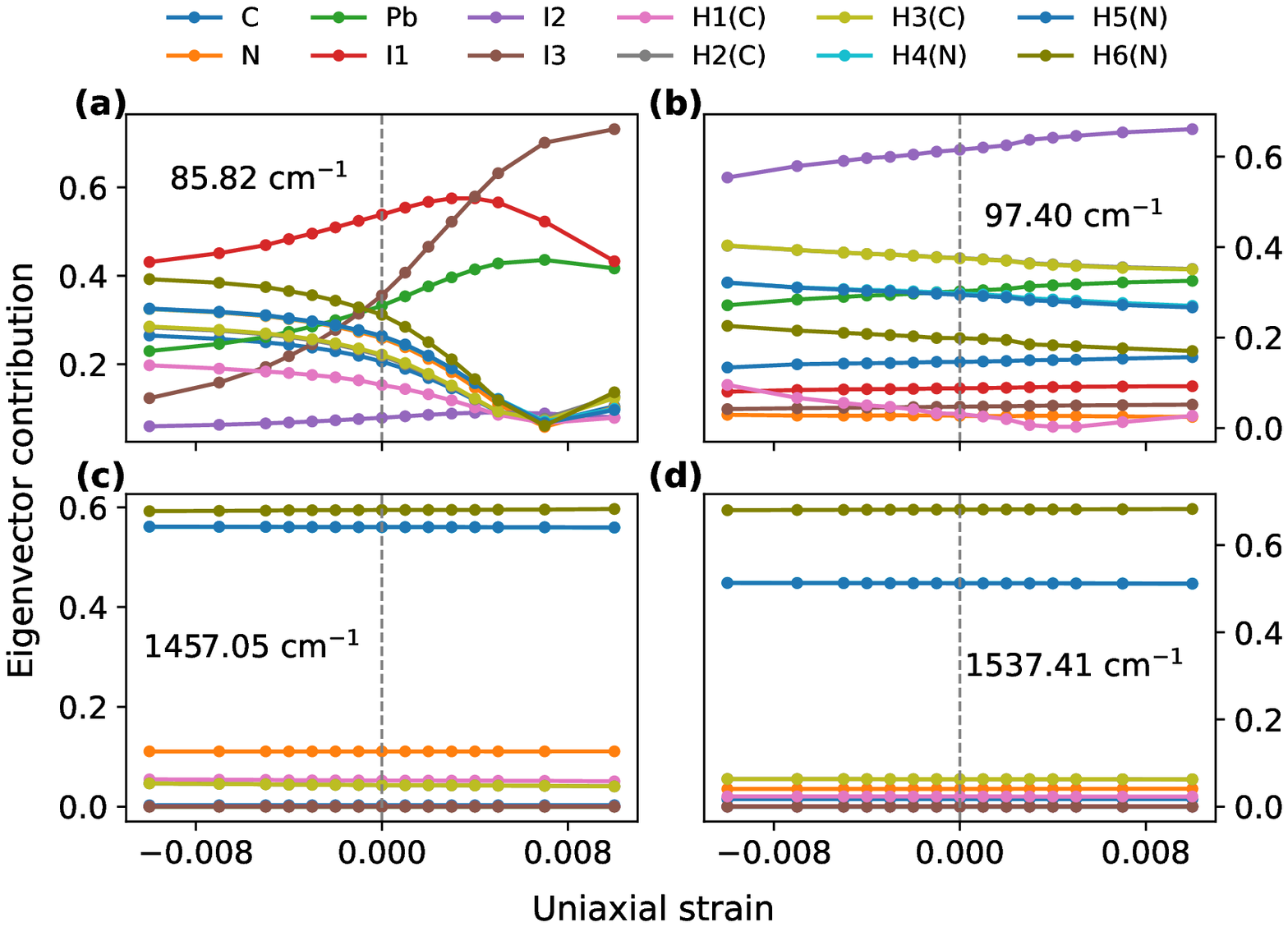}
  \caption{Mode eigenvector \textit{vs.} strain for the best modes for IR/Raman microscopy to probe local strain, showing little change in mode character (except for (a)), associated with robust linear changes \textit{vs.} strain.}
  \label{fig:bestRamanmode_eigenvector}
\end{figure}

An appropriate mode for strain mapping with IR/Raman microscopy should have \rm{i}) a large IR or Raman intensity, for ease of detection; \rm{ii}) a linear frequency change with strain, for simple and unique relation of frequency to strain; \rm{iii}) a large slope for frequency \textit{vs.} strain, for sufficient signal-to-noise ratio in measuring frequency shifts; and finally, \rm{iv}) a small change in frequency with respect to cation orientation, ensuring the validity of our results in the presence of cation rotations at elevated temperature.\citep{even2015pedestrian}\par{}

A difficulty in checking these frequency changes with cation orientation is that phonon calculations are not very meaningful except from a relaxed structure, and only a few cation orientations are stable.\citep{Brivio,Whalley2017} We took our zero-strain structure and rotated the cation close to the [111] direction, and relaxed the structure. The result is 0.01 eV higher in energy, and has the cation in the $xz$ plane, with the C-N bond at an angle of $113.9^\circ$ (compared to $23.3^\circ$ for our main data set) with respect to [100] direction. From a phonon calculation, we find some mode frequency changes due to the CH$_3$NH$_3^+$ rotation, typically by a few wavenumbers. We may expect this variation to be a contribution to heterogeneous broadening of Raman peaks as ions rotate thermally. Indeed Nakada \textit{et al.} report larger peak widths for high- and low-frequency modes than for mid-frequency modes, for all 3 crystal phases,\citep{nakada2019temperature} in accordance with our results for frequency changes with ion rotation. The high- and low-frequency modes change most while the mid-frequency modes change less due to the CH$_3$NH$_3^+$ rotation. This is because the mid-frequency region is mainly symmetric/asymmetric bending modes of the MA ion which are not affected much by rotation of the MA ion, whereas stretching modes at high and low frequency are the ones that change more when the distances change between the MA ion and the Pb-I cage. For this reason, the same modes that have large frequency changes with ion rotation have large slopes with strain. Full results are shown in Table S4. If the mode frequency changes by a significant amount due to rotation of the molecule, our static result may have a larger error, and the predicted frequency shifts due to strain will be less reliable. Therefore, we need to focus on those modes for which frequency change due to rotation of the MA ion is minimal.\par{}
After this analysis, we can identify the best modes for IR or Raman microscopy for probing local strain.\citep{de1996micro,Strubbe} We note 4 suitable modes, whose properties are detailed in Table \ref{tab:bestIR_Raman} and whose displacement patterns are shown in Fig. S13. A combination of all 4 modes can be used together for better precision, or even to find a best fit to 3 directions of uniaxial strain. Two are low-frequency Pb-I modes and two are mid-frequency molecular modes. The frequency changes \textit{vs.} strain are shown in Fig. \ref{fig:bestRamanmode_freq} and the eigenvectors \textit{vs.} strain are shown in Fig. \ref{fig:bestRamanmode_eigenvector}, exhibiting little change in mode character and a robust linear frequency change. Each of these modes has all slopes positive or all slopes negative. In the case of triaxial strain or when cation rotations wash out directional dependence, the average of the three uniaxial slopes would be the appropriate slope according to Eq. \ref{changeFreq_eps2}. Having all uniaxial slopes with the same sign is convenient, because the three directions will reinforce each other rather than cancelling out, as could happen when the slopes have different signs. The IR and Raman intensity for these 4 modes show only moderate changes with strain (Fig. S16-17) as for c-Si optical modes,\cite{de1996micro} which we expect would not cause any problem for experimental measurement.
We find another mode at 1365.3 cm$^{-1}$ that has favorable properties with reasonable Raman intensity to probe local strain (Fig. S12); however it is not observed in experiment. The reason may be limitations of our model. Our structure is pseudocubic rather than cubic, and our calculation is static and done at 0 K at which this is not the stable phase. As a result, this mode may in fact lose its Raman activity due to dynamic average symmetry in the real high-temperature structure.\par{} 
There is a final point to consider in assessing experimental feasibility. To be able to measure strain by Raman shifts, we need to obtain a frequency shift that is higher than the experimental resolution. For typical strain\citep{faghihnasiri2017dft,zhang2018strain} of 1\% we can expect to obtain shifts -3.8 cm$^{-1}$, -3.5 cm$^{-1}$, 0.75 cm$^{-1}$ and 1.57 cm$^{-1}$ for the favorable Raman modes at 85.8 cm$^{-1}$, 97.4 cm$^{-1}$, 1457.1 cm$^{-1}$ and 1537.4 cm$^{-1}$, respectively based on the average slopes in Table \ref{tab:bestIR_Raman}.
Experimentally measured full widths at half maximum (FWHMs) for Raman spectra\citep{qiu2019room,Luan,nakada2019temperature} around the low-frequency modes are  8 cm$^{-1}$, 10 cm$^{-1}$, and 18 cm$^{-1}$, and for the mode at 1460 cm$^{-1}$, the FWHM is around 35 cm$^{-1}$. One contribution to FWHM can be strain itself -- inhomogeneous strain distributions within the Raman focus will result in broadening.\citep{wu2018giant} Although these FWHMs are high, peak shifts can be resolved less than 1 cm$^{-1}$, as seen for the temperature-dependent Raman shifts measured by Nakada {\it et al.}\citep{nakada2019temperature} and in Strubbe \textit{et al.}\citep{Strubbe} on a-Si:H, because fitting of Gaussian peaks allows determination of differences in peak centers to much greater precision than the FWHM. 
Temperature dependent frequency shifts were measured for MAPI by  Nakada {\it et al.}\citep{nakada2019temperature} and they were able to measure shifts less than 1 cm$^{-1}$ which is well within the expected shifts from 1\% strain in case of our cubic structure. One contribution to the temperature-dependent frequency shift is thermal expansion which is a sign of significant effect of anharmonicity. Our results show that this is a relatively minor contribution because our predicted shifts due to thermal expansion are much less than the shifts reported by Nakada {\it et al.},\citep{nakada2019temperature} in accordance with
results of Bonini {\it et al.} for graphite.\citep{bonini2007phonon}

Finally, note that the deviation between calculated and measured frequencies for these modes is larger than expected strain shifts. Due to this unavoidable systematic difference, we would recommend use of our calculated slopes to infer relative strains in a sample, rather than by direct comparison of experimental frequencies to our calculated frequencies.\citep{Strubbe}

\subsection{Calculation of mode Gr{\"u}neisen parameter}
\begin{figure}[h!]
\centering
  \includegraphics[width=\linewidth]{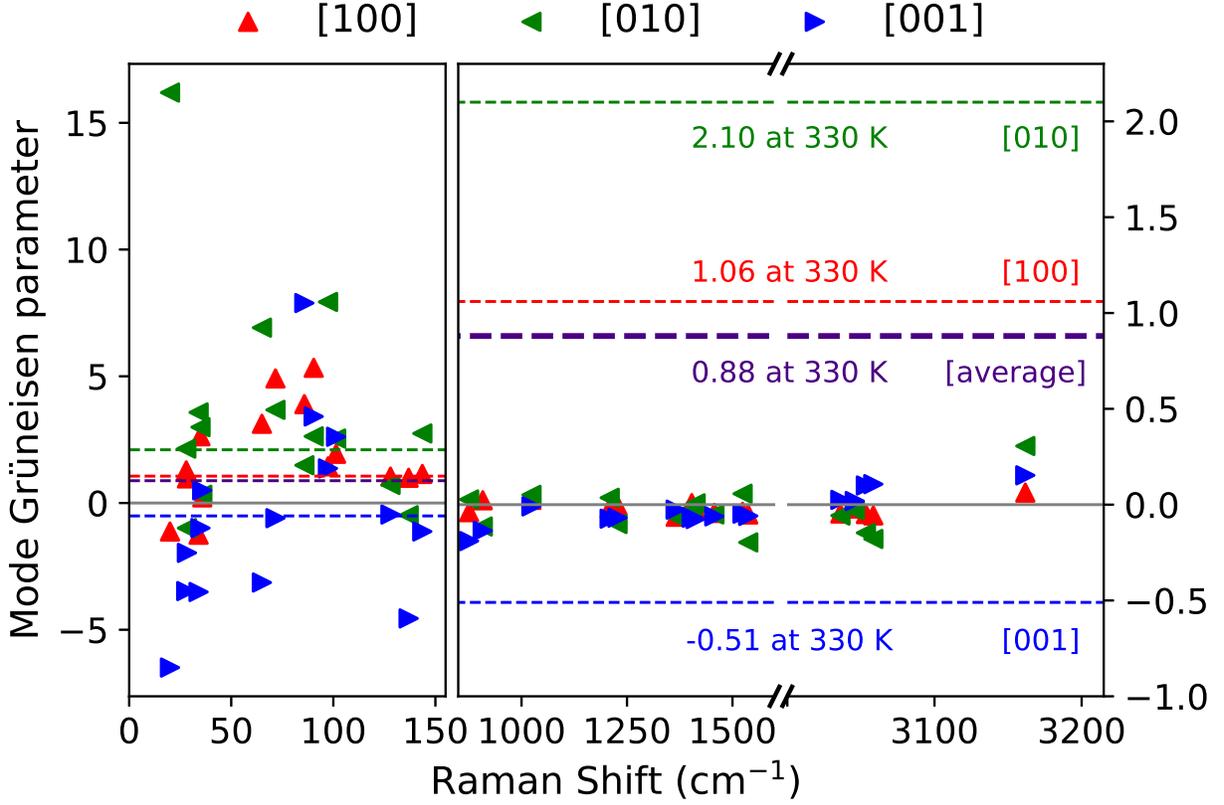}
  \caption{Calculated uniaxial mode Gr\"uneisen parameters using slopes of the frequency vs strain graph for each mode in three different crystallographic directions. Dashed lines in red, green and blue represent Gr\"uneisen parameter calculated at 330 K for different uniaxial directions. Thick dashed line in violet represents the average Gr\"uneisen parameter.}
  \label{fig:GP}
\end{figure}

\begin{table*}[t]
\setlength{\tabcolsep}{2pt}
\caption{Best modes for IR or Raman microscopy to probe local strain.\textsuperscript{\emph{a}}}
\label{tab:bestIR_Raman}
\begin{tabular}{rrrrrrrrr}
\hhline{=========}           
 \textbf{\begin{tabular}{@{}r@{}}Calc.\\ Freq.\\ (cm$^{-1}$)\end{tabular}}&\textbf{\begin{tabular}{@{}r@{}}Expt.\\ Freq.\\ (cm$^{-1}$)\end{tabular}}&\textbf{\begin{tabular}{@{}r@{}}IR\\ Int.\end{tabular}}&\textbf{\begin{tabular}{@{}r@{}}Raman\\ Int.\end{tabular}}&\textbf{\begin{tabular}{@{}r@{}}Avg.\\slope\\(cm$^{-1}$)\end{tabular}}&\textbf{\begin{tabular}{@{}r@{}}Slope\\ {[}100{]}\\(cm$^{-1}$)\end{tabular}}&\textbf{\begin{tabular}{@{}r@{}}Slope\\ {[}010{]}\\(cm$^{-1}$)\end{tabular}}&\textbf{\begin{tabular}{@{}r@{}}Slope\\ {[}001{]}\\(cm$^{-1}$)\end{tabular}}&\textbf{\begin{tabular}{@{}r@{}}Mode\\ Characters\end{tabular}}\\
 \hline

85.8&86.68\citep{qiu2019room}&	4.77&	193.2&-379.87&	-334.8&	-127.8&	-677.0&
           \begin{tabular}{@{}r@{}}CH$_3$NH$_3^+$ libration\\ CH$_3$NH$_3^+$ translation\\Pb-I-Pb bending\\Pb-I asym. stretch\end{tabular}\\ \hline
           
97.4&\begin{tabular}{@{}r@{}}94.0\citep{Luan}\\98.0\citep{nakada2019temperature} \end{tabular}&	5.53&	20.2&-348.77&	-140.8&	-772.7&	-132.8&
          \begin{tabular}{@{}r@{}}CH$_3$NH$_3^+$ libration, spin\\ Pb-I-Pb bending \end{tabular} \\ \hline           
            
1457.1&1460.0\citep{nakada2019temperature}&	5.65&	138.9&75.5&	63.2&	74.8&	88.5& 
        \begin{tabular}{@{}r@{}}C-N stretch\\N-H sym. bending\\C-H asym. bending\\no Pb-I vibration\end{tabular}\\ \hline    
        
1537.4&1582.0\citep{nakada2019temperature}&	7.48	&15.6&157.33&	79.2&	302.3&	90.5&
         \begin{tabular}{@{}r@{}}C-H asym. bending\\ N-H asym. bending\\no Pb-I vibration\end{tabular}\\
                
\hhline{=========}

\end{tabular}
\textsuperscript{\emph{a}}Absolute IR intensity is in (D/\AA)$^2$/amu and Raman intensity in \AA$^4$/amu.
\end{table*}
 The uniaxial mode Gr{\"u}neisen parameter is calculated using the slope of the frequency vs strain curve for each mode (Fig. \ref{fig:GP}), as done in a-Si.\citep{Strubbe} Pb-I vibrations at lower frequency have significant values, whereas CH$_3$NH$_3^+$ ion vibrations at higher frequency have much smaller values. One of the reasons for the low values at high frequency is the high frequency itself, as we divide the slope with the mode frequency, although some high-frequency modes do have large absolute slopes. To connect to macroscopic properties, we calculate the Gr{\"u}neisen parameter, as the weighted average over all the modes using the Bose-Einstein formula for heat capacity. In distinction to F. Brivio {\it et al.},\citep{Brivio} we are using uniaxial strain, and we include only modes at $q=\Gamma$, because of the conceptual problem in the quasiharmonic approximation of how to handle the imaginary frequencies.\cite{Deringer} Also, due to dynamical disorder \citep{leguy2016dynamic}, which is not included in our calculation, the phonon bandstructure away from $q=\Gamma$ may be less accurate or lose some of its meaning without periodicity. The imaginary frequencies are indicative of the fact that the cubic structure is not the most stable phase at $T=0$, and were observed in previous work.\citep{Brivio} They occur around $q$=R and $q$=M, and are not reduced by strain (Table S6).
Our calculated values of the directional Gr{\"u}neisen parameter at 330 K (the transition temperature to tetragonal) for strain along [100], [010], and [001] are 1.06, 2.10, and -0.51, respectively, for an average of 0.88. The isotropic value reported by F. Brivio {\it et al.},\citep{Brivio} averaged over temperature, was 1.6. We expect the difference is due to the handling of Brillouin zone integration, as well as their calculation via quasiharmonic thermal expansion rather than mode Gr{\"u}neisen parameters. This result shows a directional dependency of the Gr{\"u}neisen parameter that suggests an unusual negative thermal expansion along [001] direction for cubic MAPI, due to the negative value. While the existence of cation rotations at 330 K complicates the physical interpretation of this result, it connects to previous studies reporting that tetragonal perovskites have a negative thermal expansion coefficient along the [001] or $c$-axis,\citep{Chunyu,heiderhoff2017thermal} which is perpendicular to the direction of polarization of the CH$_3$NH$_3^+$ ion and also the largest lattice vector, in both cubic and tetragonal structures.


\section{Conclusion}
We have comprehensively studied the structural and vibrational properties of cubic CH$_3$NH$_3$PbI$_3$ under uniaxial strain. By analysing the dynamical matrix of the system under each strain we are able to identify the interactions which contribute significantly to the frequency changes due to applied strain. The phonon modes have frequency changes under strain, which can show linear, parabolic, or irregular patterns. Linear is associated with change in dynamical matrix but no change in mode eigenvector with strain, while parabolic is associated with changes in both dynamical matrix and mode eigenvector, in accordance with a perturbative analysis. Irregular patterns are due to large changes in mode eigenvector due to discontinuous structural changes, which are a sign of anharmonicity. These changes give insight into the interplay between structure, strain, and vibrations, and show approximate symmetries for some modes. We find that not only the Pb-I bond lengths but also the Pb-I-Pb bond angles change under strain, showing a buckling of the Pb-I lattice. Decrease in N-H bond length under compressive strain may be connected to the increased stability under compressive strain. We also find that the cation rotates with respect to the lattice under strain. We have identified 4 modes that are promising for IR or Raman microscopy measurement of local strain, as done in other semiconductors and even amorphous Si which has broad peaks. Results on the mode Gr\"uneisen parameters and macroscopic Gr\"uneisen parameters give insight into anharmonicity and directional negative thermal expansion. Our study of the cubic phase is relevant not only to high temperatures but also to cubic phases stabilized by ligands or other cations. Our results are expected to be similar in many respects to results for the tetragonal and orthorhombic phases, particularly in the mid-frequency range where the spectra are similar, and these phases will be the subject of future publications. We look forward to experimental work to confirm these results with uniform strain, and employ our calibration for strain mapping. Our work opens the for a standard bench-top characterization method to be usable for analyzing the critical role of local strain in hybrid perovskite photovoltaics.

\begin{suppinfo}
Structural parameters, bandstructure and bandgap, phonon mode
characterization, structural changes with strain, mode eigenvector and
frequency changes with strain, imaginary modes at the $q$=M and $q$=R points (PDF).
Relaxed atomic coordinates at each strain including phonon eigenvectors (AXSF).
\end{suppinfo}
  
\begin{acknowledgement}
We acknowledge a helpful discussion with Yong Zhang. Work was supported by UC Merced start-up funds and the Merced nAnomaterials Center for Energy and Sensing (MACES), a NASA-funded research and education center, under award NNX15AQ01. This work used computational resources from the Multi-Environment Computer for Exploration and Discovery (MERCED) cluster at UC Merced, funded by National Science Foundation Grant No. ACI-1429783, and the National Energy Research Scientific Computing Center (NERSC), a U.S. Department of Energy Office of Science User Facility operated under Contract No. DE-AC02-05CH11231.
\end{acknowledgement}

\providecommand{\latin}[1]{#1}
\makeatletter
\providecommand{\doi}
  {\begingroup\let\do\@makeother\dospecials
  \catcode`\{=1 \catcode`\}=2 \doi@aux}
\providecommand{\doi@aux}[1]{\endgroup\texttt{#1}}
\makeatother
\providecommand*\mcitethebibliography{\thebibliography}
\csname @ifundefined\endcsname{endmcitethebibliography}
  {\let\endmcitethebibliography\endthebibliography}{}


\section*{TOC Graphic}
\label{For Table of Contents Only}
    \centering
    \includegraphics[]{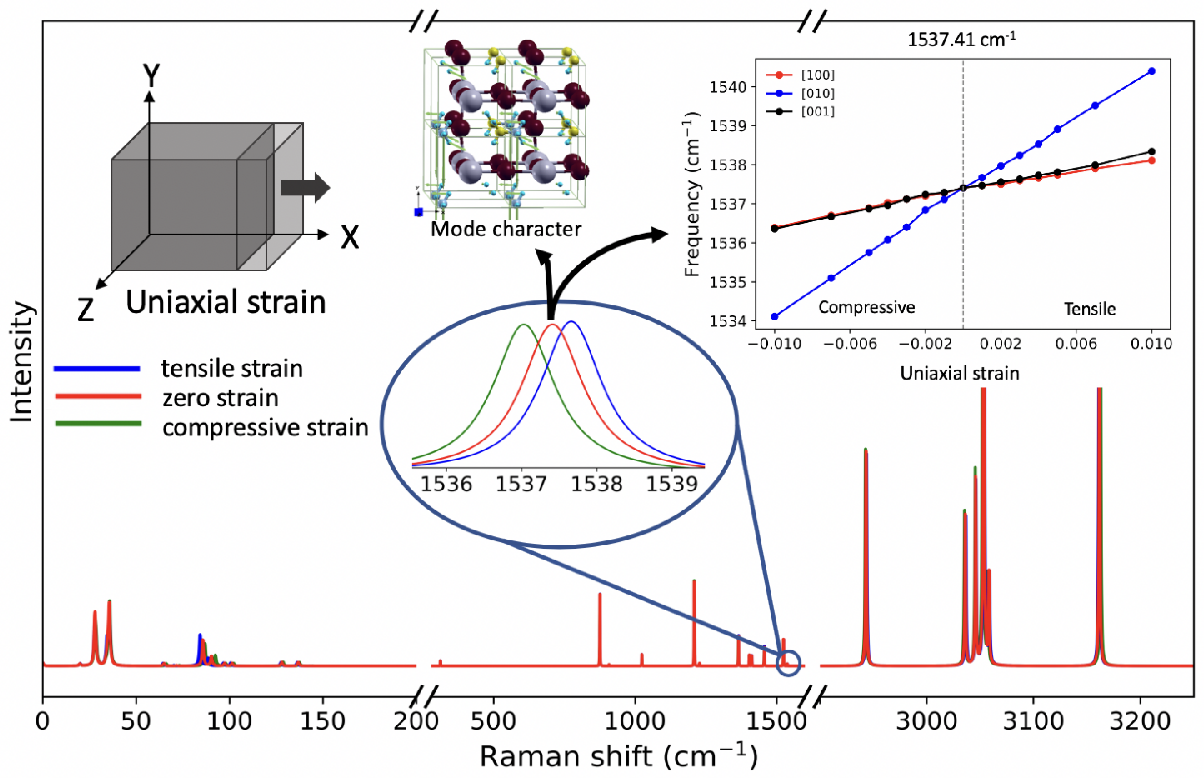}


\begin{mcitethebibliography}{99}
\providecommand*\natexlab[1]{#1}
\providecommand*\mciteSetBstSublistMode[1]{}
\providecommand*\mciteSetBstMaxWidthForm[2]{}
\providecommand*\mciteBstWouldAddEndPuncttrue
  {\def\EndOfBibitem{\unskip.}}
\providecommand*\mciteBstWouldAddEndPunctfalse
  {\let\EndOfBibitem\relax}
\providecommand*\mciteSetBstMidEndSepPunct[3]{}
\providecommand*\mciteSetBstSublistLabelBeginEnd[3]{}
\providecommand*\EndOfBibitem{}
\mciteSetBstSublistMode{f}
\mciteSetBstMaxWidthForm{subitem}{(\alph{mcitesubitemcount})}
\mciteSetBstSublistLabelBeginEnd
  {\mcitemaxwidthsubitemform\space}
  {\relax}
  {\relax}

\bibitem[Green \latin{et~al.}(2014)Green, Ho-Baillie, and Snaith]{green}
Green,~M.~A.; Ho-Baillie,~A.; Snaith,~H.~J. The emergence of perovskite solar
  cells. \emph{Nat. Photon.} \textbf{2014}, \emph{8}, 506--514\relax
\mciteBstWouldAddEndPuncttrue
\mciteSetBstMidEndSepPunct{\mcitedefaultmidpunct}
{\mcitedefaultendpunct}{\mcitedefaultseppunct}\relax
\EndOfBibitem
\bibitem[Frohna \latin{et~al.}(2018)Frohna, Deshpande, Harter, Peng, Barker,
  Neaton, Louie, Bakr, Hsieh, and Bernardi]{Frohna2018}
Frohna,~K.; Deshpande,~T.; Harter,~J.; Peng,~W.; Barker,~B.~A.; Neaton,~J.~B.;
  Louie,~S.~G.; Bakr,~O.~M.; Hsieh,~D.; Bernardi,~M. Inversion symmetry and
  bulk Rashba effect in methylammonium lead iodide perovskite single crystals.
  \emph{Nat. Commun.} \textbf{2018}, \emph{9}, 1829\relax
\mciteBstWouldAddEndPuncttrue
\mciteSetBstMidEndSepPunct{\mcitedefaultmidpunct}
{\mcitedefaultendpunct}{\mcitedefaultseppunct}\relax
\EndOfBibitem
\bibitem[Umari \latin{et~al.}(2014)Umari, Mosconi, and
  De~Angelis]{Umari2014RelativisticGC}
Umari,~P.; Mosconi,~E.; De~Angelis,~F. Relativistic GW calculations on
  CH$_3$NH$_3$PbI$_3$ and CH$_3$NH$_3$SnI$_3$ perovskites for solar cell
  applications. \emph{Sci. Rep.} \textbf{2014}, \emph{4}, 4467\relax
\mciteBstWouldAddEndPuncttrue
\mciteSetBstMidEndSepPunct{\mcitedefaultmidpunct}
{\mcitedefaultendpunct}{\mcitedefaultseppunct}\relax
\EndOfBibitem
\bibitem[Shirayama \latin{et~al.}(2016)Shirayama, Kadowaki, Miyadera, Sugita,
  Tamakoshi, Kato, Fujiseki, Murata, Hara, Murakami, Fujimoto, Chikamatsu, and
  Fujiwara]{Shirayama2016a}
Shirayama,~M.; Kadowaki,~H.; Miyadera,~T.; Sugita,~T.; Tamakoshi,~M.; Kato,~M.;
  Fujiseki,~T.; Murata,~D.; Hara,~S.; Murakami,~T.~N. \latin{et~al.}  {Optical
  transitions in hybrid perovskite solar cells: Ellipsometry, density
  functional theory, and quantum efficiency analyses for CH$_3$NH$_3$PbI$_3$}.
  \emph{Phys. Rev. Appl.} \textbf{2016}, \emph{5}, 014012\relax
\mciteBstWouldAddEndPuncttrue
\mciteSetBstMidEndSepPunct{\mcitedefaultmidpunct}
{\mcitedefaultendpunct}{\mcitedefaultseppunct}\relax
\EndOfBibitem
\bibitem[Stranks \latin{et~al.}(2013)Stranks, Eperon, Grancini, Menelaou,
  Alcocer, Leijtens, Herz, Petrozza, and Snaith]{Stranks341}
Stranks,~S.~D.; Eperon,~G.~E.; Grancini,~G.; Menelaou,~C.; Alcocer,~M. J.~P.;
  Leijtens,~T.; Herz,~L.~M.; Petrozza,~A.; Snaith,~H.~J. Electron-hole
  diffusion lengths exceeding 1 micrometer in an organometal trihalide
  perovskite absorber. \emph{Science} \textbf{2013}, \emph{342}, 341--344\relax
\mciteBstWouldAddEndPuncttrue
\mciteSetBstMidEndSepPunct{\mcitedefaultmidpunct}
{\mcitedefaultendpunct}{\mcitedefaultseppunct}\relax
\EndOfBibitem
\bibitem[Xing \latin{et~al.}(2013)Xing, Mathews, Sun, Lim, Lam, Gr{\"a}tzel,
  Mhaisalkar, and Sum]{Xing344}
Xing,~G.; Mathews,~N.; Sun,~S.; Lim,~S.~S.; Lam,~Y.~M.; Gr{\"a}tzel,~M.;
  Mhaisalkar,~S.; Sum,~T.~C. Long-range balanced electron- and hole-transport
  lengths in organic-inorganic CH$_3$NH$_3$PbI$_3$. \emph{Science}
  \textbf{2013}, \emph{342}, 344--347\relax
\mciteBstWouldAddEndPuncttrue
\mciteSetBstMidEndSepPunct{\mcitedefaultmidpunct}
{\mcitedefaultendpunct}{\mcitedefaultseppunct}\relax
\EndOfBibitem
\bibitem[Wehrenfennig \latin{et~al.}(2014)Wehrenfennig, Eperon, Johnston,
  Snaith, and Herz]{Wehrenfennig2014}
Wehrenfennig,~C.; Eperon,~G.~E.; Johnston,~M.~B.; Snaith,~H.~J.; Herz,~L.~M.
  High charge carrier mobilities and lifetimes in organolead trihalide
  perovskites. \emph{Adv. Mater.} \textbf{2014}, \emph{26}, 1584--1589\relax
\mciteBstWouldAddEndPuncttrue
\mciteSetBstMidEndSepPunct{\mcitedefaultmidpunct}
{\mcitedefaultendpunct}{\mcitedefaultseppunct}\relax
\EndOfBibitem
\bibitem[Kojima \latin{et~al.}(2009)Kojima, Teshima, Shirai, and
  Miyasaka]{Kojima2009}
Kojima,~A.; Teshima,~K.; Shirai,~Y.; Miyasaka,~T. Organometal halide
  perovskites as visible-light sensitizers for photovoltaic cells. \emph{J. Am.
  Chem. Soc} \textbf{2009}, \emph{131}, 6050--6051\relax
\mciteBstWouldAddEndPuncttrue
\mciteSetBstMidEndSepPunct{\mcitedefaultmidpunct}
{\mcitedefaultendpunct}{\mcitedefaultseppunct}\relax
\EndOfBibitem
\bibitem[NRE()]{NREL}
Best Research-Cell Efficiency Chart from National Renewable Energy Laboratory.
  \url{https://www.nrel.gov/pv/assets/pdfs/best-research-cell-efficiencies.20200708.pdf}
  (accessed September 27, 2020)\relax
\mciteBstWouldAddEndPuncttrue
\mciteSetBstMidEndSepPunct{\mcitedefaultmidpunct}
{\mcitedefaultendpunct}{\mcitedefaultseppunct}\relax
\EndOfBibitem
\bibitem[Eperon \latin{et~al.}(2016)Eperon, Leijtens, Bush, Prasanna, Green,
  Wang, McMeekin, Volonakis, Milot, May, Palmstrom, Slotcavage, Belisle, Patel,
  Parrott, Sutton, Ma, Moghadam, Conings, Babayigit, Boyen, Bent, Giustino,
  Herz, Johnston, McGehee, and Snaith]{Eperon861}
Eperon,~G.~E.; Leijtens,~T.; Bush,~K.~A.; Prasanna,~R.; Green,~T.; Wang,~J.
  T.-W.; McMeekin,~D.~P.; Volonakis,~G.; Milot,~R.~L.; May,~R. \latin{et~al.}
  Perovskite-perovskite tandem photovoltaics with optimized band gaps.
  \emph{Science} \textbf{2016}, \emph{354}, 861--865\relax
\mciteBstWouldAddEndPuncttrue
\mciteSetBstMidEndSepPunct{\mcitedefaultmidpunct}
{\mcitedefaultendpunct}{\mcitedefaultseppunct}\relax
\EndOfBibitem
\bibitem[Lin \latin{et~al.}(2015)Lin, Armin, Nagiri, Burn, and
  Meredith]{Lin2014}
Lin,~Q.; Armin,~A.; Nagiri,~R. C.~R.; Burn,~P.~L.; Meredith,~P. Electro-optics
  of perovskite solar cells. \emph{Nat. Photon.} \textbf{2015}, \emph{9},
  106\relax
\mciteBstWouldAddEndPuncttrue
\mciteSetBstMidEndSepPunct{\mcitedefaultmidpunct}
{\mcitedefaultendpunct}{\mcitedefaultseppunct}\relax
\EndOfBibitem
\bibitem[Yang \latin{et~al.}(2019)Yang, Yang, Priya, and Liu]{FlexiblePVSK}
Yang,~D.; Yang,~R.; Priya,~S.; Liu,~S.~F. Recent advances in flexible
  perovskite solar cells: Fabrication and applications. \emph{Angew. Chem. Int.
  Ed.} \textbf{2019}, \emph{58}, 4466--4483\relax
\mciteBstWouldAddEndPuncttrue
\mciteSetBstMidEndSepPunct{\mcitedefaultmidpunct}
{\mcitedefaultendpunct}{\mcitedefaultseppunct}\relax
\EndOfBibitem
\bibitem[Bailie \latin{et~al.}(2015)Bailie, Christoforo, Mailoa, Bowring,
  Unger, Nguyen, Burschka, Pellet, Lee, Gr{\"a}tzel, \latin{et~al.}
  others]{bailie2015semi}
Bailie,~C.~D.; Christoforo,~M.~G.; Mailoa,~J.~P.; Bowring,~A.~R.; Unger,~E.~L.;
  Nguyen,~W.~H.; Burschka,~J.; Pellet,~N.; Lee,~J.~Z.; Gr{\"a}tzel,~M.
  \latin{et~al.}  Semi-transparent perovskite solar cells for tandems with
  silicon and CIGS. \emph{Energy Environ. Sci} \textbf{2015}, \emph{8},
  956--963\relax
\mciteBstWouldAddEndPuncttrue
\mciteSetBstMidEndSepPunct{\mcitedefaultmidpunct}
{\mcitedefaultendpunct}{\mcitedefaultseppunct}\relax
\EndOfBibitem
\bibitem[Rold{\'a}n-Carmona \latin{et~al.}(2014)Rold{\'a}n-Carmona,
  Malinkiewicz, Soriano, Espallargas, Garcia, Reinecke, Kroyer, Dar,
  Nazeeruddin, and Bolink]{roldan2014flexible}
Rold{\'a}n-Carmona,~C.; Malinkiewicz,~O.; Soriano,~A.; Espallargas,~G.~M.;
  Garcia,~A.; Reinecke,~P.; Kroyer,~T.; Dar,~M.~I.; Nazeeruddin,~M.~K.;
  Bolink,~H.~J. Flexible high efficiency perovskite solar cells. \emph{Energy
  Environ. Sci} \textbf{2014}, \emph{7}, 994--997\relax
\mciteBstWouldAddEndPuncttrue
\mciteSetBstMidEndSepPunct{\mcitedefaultmidpunct}
{\mcitedefaultendpunct}{\mcitedefaultseppunct}\relax
\EndOfBibitem
\bibitem[Ping and Zhang(2018)Ping, and Zhang]{ping2018spin}
Ping,~Y.; Zhang,~J.~Z. Spin-optotronic properties of organometal halide
  perovskites. \emph{J. Phys. Chem. Lett.} \textbf{2018}, \emph{9},
  6103--6111\relax
\mciteBstWouldAddEndPuncttrue
\mciteSetBstMidEndSepPunct{\mcitedefaultmidpunct}
{\mcitedefaultendpunct}{\mcitedefaultseppunct}\relax
\EndOfBibitem
\bibitem[Nikolaidou \latin{et~al.}(2016)Nikolaidou, Sarang, Hoffman, Mendewala,
  Ishihara, Lu, Ilan, Tung, and Ghosh]{nikolaidou2016hybrid}
Nikolaidou,~K.; Sarang,~S.; Hoffman,~C.; Mendewala,~B.; Ishihara,~H.;
  Lu,~J.~Q.; Ilan,~B.; Tung,~V.; Ghosh,~S. Hybrid perovskite thin films as
  highly efficient luminescent solar concentrators. \emph{Adv. Opt. Mater.}
  \textbf{2016}, \emph{4}, 2126--2132\relax
\mciteBstWouldAddEndPuncttrue
\mciteSetBstMidEndSepPunct{\mcitedefaultmidpunct}
{\mcitedefaultendpunct}{\mcitedefaultseppunct}\relax
\EndOfBibitem
\bibitem[Kim \latin{et~al.}(2018)Kim, Zhao, Price, Grede, Roh, Brigeman, Lopez,
  Rand, and Giebink]{hkim}
Kim,~H.; Zhao,~L.; Price,~J.~S.; Grede,~A.~J.; Roh,~K.; Brigeman,~A.~N.;
  Lopez,~M.; Rand,~B.~P.; Giebink,~N.~C. Hybrid perovskite light-emitting
  diodes under intense electrical excitation. \emph{Nat. Commun.}
  \textbf{2018}, \emph{9}, 4893\relax
\mciteBstWouldAddEndPuncttrue
\mciteSetBstMidEndSepPunct{\mcitedefaultmidpunct}
{\mcitedefaultendpunct}{\mcitedefaultseppunct}\relax
\EndOfBibitem
\bibitem[Niu \latin{et~al.}(2015)Niu, Guo, and Wang]{niu2015review}
Niu,~G.; Guo,~X.; Wang,~L. Review of recent progress in chemical stability of
  perovskite solar cells. \emph{J. Mater. Chem. A} \textbf{2015}, \emph{3},
  8970--8980\relax
\mciteBstWouldAddEndPuncttrue
\mciteSetBstMidEndSepPunct{\mcitedefaultmidpunct}
{\mcitedefaultendpunct}{\mcitedefaultseppunct}\relax
\EndOfBibitem
\bibitem[Conings \latin{et~al.}(2015)Conings, Drijkoningen, Gauquelin,
  Babayigit, D'Haen, D'Olieslaeger, Ethirajan, Verbeeck, Manca, Mosconi,
  \latin{et~al.} others]{conings2015intrinsic}
Conings,~B.; Drijkoningen,~J.; Gauquelin,~N.; Babayigit,~A.; D'Haen,~J.;
  D'Olieslaeger,~L.; Ethirajan,~A.; Verbeeck,~J.; Manca,~J.; Mosconi,~E.
  \latin{et~al.}  Intrinsic thermal instability of methylammonium lead
  trihalide perovskite. \emph{Adv. Energy Mater.} \textbf{2015}, \emph{5},
  1500477\relax
\mciteBstWouldAddEndPuncttrue
\mciteSetBstMidEndSepPunct{\mcitedefaultmidpunct}
{\mcitedefaultendpunct}{\mcitedefaultseppunct}\relax
\EndOfBibitem
\bibitem[Joshi \latin{et~al.}(2016)Joshi, Zhang, Hossain, Abbas, Kottokkaran,
  Nehra, Dhaka, Noack, and Dalal]{Joshi2016}
Joshi,~P.~H.; Zhang,~L.; Hossain,~I.~M.; Abbas,~H.~A.; Kottokkaran,~R.;
  Nehra,~S.~P.; Dhaka,~M.; Noack,~M.; Dalal,~V.~L. The physics of photon
  induced degradation of perovskite solar cells. \emph{AIP Adv.}
  \textbf{2016}, \emph{6}, 115114\relax
\mciteBstWouldAddEndPuncttrue
\mciteSetBstMidEndSepPunct{\mcitedefaultmidpunct}
{\mcitedefaultendpunct}{\mcitedefaultseppunct}\relax
\EndOfBibitem
\bibitem[Lee \latin{et~al.}(2015)Lee, Kim, Kim, Seo, Cho, and
  Park]{lee2015formamidinium}
Lee,~J.-W.; Kim,~D.-H.; Kim,~H.-S.; Seo,~S.-W.; Cho,~S.~M.; Park,~N.-G.
  Formamidinium and cesium hybridization for photo-and moisture-stable
  perovskite solar cell. \emph{Adv. Energy Mater.} \textbf{2015}, \emph{5},
  1501310\relax
\mciteBstWouldAddEndPuncttrue
\mciteSetBstMidEndSepPunct{\mcitedefaultmidpunct}
{\mcitedefaultendpunct}{\mcitedefaultseppunct}\relax
\EndOfBibitem
\bibitem[Li \latin{et~al.}(2015)Li, Tschumi, Han, Babkair, Alzubaydi, Ansari,
  Habib, Nazeeruddin, Zakeeruddin, and Gr{\"a}tzel]{li2015outdoor}
Li,~X.; Tschumi,~M.; Han,~H.; Babkair,~S.~S.; Alzubaydi,~R.~A.; Ansari,~A.~A.;
  Habib,~S.~S.; Nazeeruddin,~M.~K.; Zakeeruddin,~S.~M.; Gr{\"a}tzel,~M. Outdoor
  performance and stability under elevated temperatures and long-term light
  soaking of triple-layer mesoporous perovskite photovoltaics. \emph{Energy
  Technol.} \textbf{2015}, \emph{3}, 551--555\relax
\mciteBstWouldAddEndPuncttrue
\mciteSetBstMidEndSepPunct{\mcitedefaultmidpunct}
{\mcitedefaultendpunct}{\mcitedefaultseppunct}\relax
\EndOfBibitem
\bibitem[Ong \latin{et~al.}(2015)Ong, Goh, Xu, and Huan]{ong2015structural}
Ong,~K.~P.; Goh,~T.~W.; Xu,~Q.; Huan,~A. Structural evolution in methylammonium
  lead iodide CH$_3$NH$_3$PbI$_3$. \emph{J. Phys. Chem. A} \textbf{2015},
  \emph{119}, 11033--11038\relax
\mciteBstWouldAddEndPuncttrue
\mciteSetBstMidEndSepPunct{\mcitedefaultmidpunct}
{\mcitedefaultendpunct}{\mcitedefaultseppunct}\relax
\EndOfBibitem
\bibitem[Nie \latin{et~al.}(2016)Nie, Blancon, Neukirch, Appavoo, Tsai,
  Chhowalla, Alam, Sfeir, Katan, Even, Tretiak, Crochet, Gupta, and
  Mohite]{Nie2016}
Nie,~W.; Blancon,~J.~C.; Neukirch,~A.~J.; Appavoo,~K.; Tsai,~H.; Chhowalla,~M.;
  Alam,~M.~A.; Sfeir,~M.~Y.; Katan,~C.; Even,~J. \latin{et~al.}
  {Light-activated photocurrent degradation and self-healing in perovskite
  solar cells}. \emph{Nat. Commun.} \textbf{2016}, \emph{7}, 1--9\relax
\mciteBstWouldAddEndPuncttrue
\mciteSetBstMidEndSepPunct{\mcitedefaultmidpunct}
{\mcitedefaultendpunct}{\mcitedefaultseppunct}\relax
\EndOfBibitem
\bibitem[Zhao \latin{et~al.}(2017)Zhao, Deng, Wei, Zheng, Yu, Shao, Shield, and
  Huang]{zhao2017strained}
Zhao,~J.; Deng,~Y.; Wei,~H.; Zheng,~X.; Yu,~Z.; Shao,~Y.; Shield,~J.~E.;
  Huang,~J. Strained hybrid perovskite thin films and their impact on the
  intrinsic stability of perovskite solar cells. \emph{Sci. Adv.}
  \textbf{2017}, \emph{3}, eaao5616\relax
\mciteBstWouldAddEndPuncttrue
\mciteSetBstMidEndSepPunct{\mcitedefaultmidpunct}
{\mcitedefaultendpunct}{\mcitedefaultseppunct}\relax
\EndOfBibitem
\bibitem[Faghihnasiri \latin{et~al.}(2017)Faghihnasiri, Izadifard, and
  Ghazi]{faghihnasiri2017dft}
Faghihnasiri,~M.; Izadifard,~M.; Ghazi,~M.~E. DFT study of mechanical
  properties and stability of cubic methylammonium lead halide perovskites
  (CH$_3$NH$_3$PbI$_3$, X= I, Br, Cl). \emph{J. Phys. Chem. C} \textbf{2017},
  \emph{121}, 27059--27070\relax
\mciteBstWouldAddEndPuncttrue
\mciteSetBstMidEndSepPunct{\mcitedefaultmidpunct}
{\mcitedefaultendpunct}{\mcitedefaultseppunct}\relax
\EndOfBibitem
\bibitem[Zhang \latin{et~al.}(2018)Zhang, Geng, Tong, Chen, Cao, and
  Chen]{zhang2018strain}
Zhang,~L.; Geng,~W.; Tong,~C.~J.; Chen,~X.; Cao,~T.; Chen,~M. Strain induced
  electronic structure variation in methyl-ammonium lead iodide perovskite.
  \emph{Sci. Rep.} \textbf{2018}, \emph{8}, 7760\relax
\mciteBstWouldAddEndPuncttrue
\mciteSetBstMidEndSepPunct{\mcitedefaultmidpunct}
{\mcitedefaultendpunct}{\mcitedefaultseppunct}\relax
\EndOfBibitem
\bibitem[Bechtel and Van~der Ven(2018)Bechtel, and Van~der
  Ven]{bechtel2018octahedral}
Bechtel,~J.~S.; Van~der Ven,~A. Octahedral tilting instabilities in inorganic
  halide perovskites. \emph{Phys. Rev. Mater.} \textbf{2018}, \emph{2},
  025401\relax
\mciteBstWouldAddEndPuncttrue
\mciteSetBstMidEndSepPunct{\mcitedefaultmidpunct}
{\mcitedefaultendpunct}{\mcitedefaultseppunct}\relax
\EndOfBibitem
\bibitem[Tsai \latin{et~al.}(2018)Tsai, Asadpour, Blancon, Stoumpos, Durand,
  Strzalka, Chen, Verduzco, Ajayan, Tretiak, \latin{et~al.} others]{tsai2018}
Tsai,~H.; Asadpour,~R.; Blancon,~J.-C.; Stoumpos,~C.~C.; Durand,~O.;
  Strzalka,~J.~W.; Chen,~B.; Verduzco,~R.; Ajayan,~P.~M.; Tretiak,~S.
  \latin{et~al.}  Light-induced lattice expansion leads to high-efficiency
  perovskite solar cells. \emph{Science} \textbf{2018}, \emph{360},
  67--70\relax
\mciteBstWouldAddEndPuncttrue
\mciteSetBstMidEndSepPunct{\mcitedefaultmidpunct}
{\mcitedefaultendpunct}{\mcitedefaultseppunct}\relax
\EndOfBibitem
\bibitem[Jones \latin{et~al.}(2019)Jones, Osherov, Alsari, Sponseller, Duck,
  Jung, Settens, Niroui, Brenes, Stan, \latin{et~al.} others]{jones2019lattice}
Jones,~T.~W.; Osherov,~A.; Alsari,~M.; Sponseller,~M.; Duck,~B.~C.;
  Jung,~Y.-K.; Settens,~C.; Niroui,~F.; Brenes,~R.; Stan,~C.~V. \latin{et~al.}
  Lattice strain causes non-radiative losses in halide perovskites.
  \emph{Energy Environ. Sci} \textbf{2019}, \emph{12}, 596--606\relax
\mciteBstWouldAddEndPuncttrue
\mciteSetBstMidEndSepPunct{\mcitedefaultmidpunct}
{\mcitedefaultendpunct}{\mcitedefaultseppunct}\relax
\EndOfBibitem
\bibitem[Zhu \latin{et~al.}(2019)Zhu, Niu, Fu, Li, Hu, Chen, He, Na, Liu, Zai,
  \latin{et~al.} others]{Zhu2019}
Zhu,~C.; Niu,~X.; Fu,~Y.; Li,~N.; Hu,~C.; Chen,~Y.; He,~X.; Na,~G.; Liu,~P.;
  Zai,~H. \latin{et~al.}  Strain engineering in perovskite solar cells and its
  impacts on carrier dynamics. \emph{Nat. Commun.} \textbf{2019}, \emph{10},
  1--11\relax
\mciteBstWouldAddEndPuncttrue
\mciteSetBstMidEndSepPunct{\mcitedefaultmidpunct}
{\mcitedefaultendpunct}{\mcitedefaultseppunct}\relax
\EndOfBibitem
\bibitem[Slotcavage \latin{et~al.}(2016)Slotcavage, Karunadasa, and
  McGehee]{slotcavage2016light}
Slotcavage,~D.~J.; Karunadasa,~H.~I.; McGehee,~M.~D. Light-induced phase
  segregation in halide-perovskite absorbers. \emph{ACS Energy Lett.}
  \textbf{2016}, \emph{1}, 1199--1205\relax
\mciteBstWouldAddEndPuncttrue
\mciteSetBstMidEndSepPunct{\mcitedefaultmidpunct}
{\mcitedefaultendpunct}{\mcitedefaultseppunct}\relax
\EndOfBibitem
\bibitem[Chen \latin{et~al.}(2020)Chen, Lei, Li, Yu, Cai, Chiu, Rao, Gu, Wang,
  Choi, \latin{et~al.} others]{chen2020strain}
Chen,~Y.; Lei,~Y.; Li,~Y.; Yu,~Y.; Cai,~J.; Chiu,~M.-H.; Rao,~R.; Gu,~Y.;
  Wang,~C.; Choi,~W. \latin{et~al.}  Strain engineering and epitaxial
  stabilization of halide perovskites. \emph{Nature} \textbf{2020}, \emph{577},
  209--215\relax
\mciteBstWouldAddEndPuncttrue
\mciteSetBstMidEndSepPunct{\mcitedefaultmidpunct}
{\mcitedefaultendpunct}{\mcitedefaultseppunct}\relax
\EndOfBibitem
\bibitem[Huang \latin{et~al.}(2017)Huang, Yuan, Shao, and
  Yan]{huang2017understanding}
Huang,~J.; Yuan,~Y.; Shao,~Y.; Yan,~Y. Understanding the physical properties of
  hybrid perovskites for photovoltaic applications. \emph{Nat. Rev. Mater.}
  \textbf{2017}, \emph{2}, 17042\relax
\mciteBstWouldAddEndPuncttrue
\mciteSetBstMidEndSepPunct{\mcitedefaultmidpunct}
{\mcitedefaultendpunct}{\mcitedefaultseppunct}\relax
\EndOfBibitem
\bibitem[Jena \latin{et~al.}(2019)Jena, Kulkarni, and Miyasaka]{jena2019halide}
Jena,~A.~K.; Kulkarni,~A.; Miyasaka,~T. Halide perovskite photovoltaics:
  background, status, and future prospects. \emph{Chem. Rev.} \textbf{2019},
  \emph{119}, 3036--3103\relax
\mciteBstWouldAddEndPuncttrue
\mciteSetBstMidEndSepPunct{\mcitedefaultmidpunct}
{\mcitedefaultendpunct}{\mcitedefaultseppunct}\relax
\EndOfBibitem
\bibitem[Even(2015)]{even2015pedestrian}
Even,~J. Pedestrian guide to symmetry properties of the reference cubic
  structure of 3D all-inorganic and hybrid perovskites. \emph{J. Phys. Chem.
  Lett.} \textbf{2015}, \emph{6}, 2238--2242\relax
\mciteBstWouldAddEndPuncttrue
\mciteSetBstMidEndSepPunct{\mcitedefaultmidpunct}
{\mcitedefaultendpunct}{\mcitedefaultseppunct}\relax
\EndOfBibitem
\bibitem[Sarang \latin{et~al.}(2017)Sarang, Bonabi~Naghadeh, Luo, Kumar,
  Betady, Tung, Scheibner, Zhang, and Ghosh]{sarang2017stabilization}
Sarang,~S.; Bonabi~Naghadeh,~S.; Luo,~B.; Kumar,~P.; Betady,~E.; Tung,~V.;
  Scheibner,~M.; Zhang,~J.~Z.; Ghosh,~S. Stabilization of the cubic crystalline
  phase in organometal halide perovskite quantum dots via surface energy
  manipulation. \emph{J. Phys. Chem. Lett.} \textbf{2017}, \emph{8},
  5378--5384\relax
\mciteBstWouldAddEndPuncttrue
\mciteSetBstMidEndSepPunct{\mcitedefaultmidpunct}
{\mcitedefaultendpunct}{\mcitedefaultseppunct}\relax
\EndOfBibitem
\bibitem[Wu \latin{et~al.}(2018)Wu, Chen, Guo, Wang, and Li]{wu2018cations}
Wu,~C.; Chen,~K.; Guo,~D.; Wang,~S.; Li,~P. Cations substitution tuning phase
  stability in hybrid perovskite single crystals by strain relaxation.
  \emph{RSC Adv.} \textbf{2018}, \emph{8}, 2900--2905\relax
\mciteBstWouldAddEndPuncttrue
\mciteSetBstMidEndSepPunct{\mcitedefaultmidpunct}
{\mcitedefaultendpunct}{\mcitedefaultseppunct}\relax
\EndOfBibitem
\bibitem[Sheng \latin{et~al.}(2015)Sheng, Ho-Baillie, Huang, Chen, Wen, Hao,
  and Green]{sheng2015methylammonium}
Sheng,~R.; Ho-Baillie,~A.; Huang,~S.; Chen,~S.; Wen,~X.; Hao,~X.; Green,~M.~A.
  Methylammonium lead bromide perovskite-based solar cells by vapor-assisted
  deposition. \emph{J. Phys. Chem. C} \textbf{2015}, \emph{119},
  3545--3549\relax
\mciteBstWouldAddEndPuncttrue
\mciteSetBstMidEndSepPunct{\mcitedefaultmidpunct}
{\mcitedefaultendpunct}{\mcitedefaultseppunct}\relax
\EndOfBibitem
\bibitem[Zhang \latin{et~al.}(2015)Zhang, Qiao, Shen, Moehl, Zakeeruddin,
  Gr{\"a}tzel, and Wang]{zhang2015photovoltaic}
Zhang,~H.; Qiao,~X.; Shen,~Y.; Moehl,~T.; Zakeeruddin,~S.~M.; Gr{\"a}tzel,~M.;
  Wang,~M. Photovoltaic behaviour of lead methylammonium triiodide perovskite
  solar cells down to 80 K. \emph{J. Mater. Chem. A} \textbf{2015}, \emph{3},
  11762--11767\relax
\mciteBstWouldAddEndPuncttrue
\mciteSetBstMidEndSepPunct{\mcitedefaultmidpunct}
{\mcitedefaultendpunct}{\mcitedefaultseppunct}\relax
\EndOfBibitem
\bibitem[Brivio \latin{et~al.}(2015)Brivio, Frost, Skelton, Jackson, Weber,
  Weller, Go\~ni, Leguy, Barnes, and Walsh]{Brivio}
Brivio,~F.; Frost,~J.~M.; Skelton,~J.~M.; Jackson,~A.~J.; Weber,~O.~J.;
  Weller,~M.~T.; Go\~ni,~A.~R.; Leguy,~A. M.~A.; Barnes,~P. R.~F.; Walsh,~A.
  Lattice dynamics and vibrational spectra of the orthorhombic, tetragonal, and
  cubic phases of methylammonium lead iodide. \emph{Phys. Rev. B}
  \textbf{2015}, \emph{92}, 144308\relax
\mciteBstWouldAddEndPuncttrue
\mciteSetBstMidEndSepPunct{\mcitedefaultmidpunct}
{\mcitedefaultendpunct}{\mcitedefaultseppunct}\relax
\EndOfBibitem
\bibitem[Luan \latin{et~al.}(2016)Luan, Song, Wei, Chen, and Liu]{Luan}
Luan,~M.; Song,~J.; Wei,~X.; Chen,~F.; Liu,~J. Controllable growth of bulk
  cubic-phase CH$_3$NH$_3$PbI$_3$ single crystal with exciting room-temperature
  stability. \emph{CrystEngComm} \textbf{2016}, \emph{18}, 5257--5261\relax
\mciteBstWouldAddEndPuncttrue
\mciteSetBstMidEndSepPunct{\mcitedefaultmidpunct}
{\mcitedefaultendpunct}{\mcitedefaultseppunct}\relax
\EndOfBibitem
\bibitem[Qiu \latin{et~al.}(2019)Qiu, McDowell, and Shi]{qiu2019room}
Qiu,~J.; McDowell,~L.~L.; Shi,~Z. Room-temperature cubic perovskite thin films
  by three-step all-vapor conversion from PbSe to MAPbI$_3$. \emph{Cryst.
  Growth Des.} \textbf{2019}, \emph{19}, 2001--2009\relax
\mciteBstWouldAddEndPuncttrue
\mciteSetBstMidEndSepPunct{\mcitedefaultmidpunct}
{\mcitedefaultendpunct}{\mcitedefaultseppunct}\relax
\EndOfBibitem
\bibitem[Ledinsk\'{y} \latin{et~al.}(2015)Ledinsk\'{y}, L\"{o}per, Niesen,
  Holovsk\'{y}, Moon, Yum, De~Wolf, Fejfar, and Ballif]{ledinsky2015raman}
Ledinsk\'{y},~M.; L\"{o}per,~P.; Niesen,~B.; Holovsk\'{y},~J.; Moon,~S.-J.;
  Yum,~J.-H.; De~Wolf,~S.; Fejfar,~A.; Ballif,~C. Raman spectroscopy of
  organic--inorganic halide perovskites. \emph{J. Phys. Chem. Lett.}
  \textbf{2015}, \emph{6}, 401--406\relax
\mciteBstWouldAddEndPuncttrue
\mciteSetBstMidEndSepPunct{\mcitedefaultmidpunct}
{\mcitedefaultendpunct}{\mcitedefaultseppunct}\relax
\EndOfBibitem
\bibitem[Nakada \latin{et~al.}(2019)Nakada, Matsumoto, Shimoi, Yamada, and
  Furukawa]{nakada2019temperature}
Nakada,~K.; Matsumoto,~Y.; Shimoi,~Y.; Yamada,~K.; Furukawa,~Y.
  Temperature-dependent evolution of Raman spectra of methylammonium lead
  halide perovskites, CH$_3$NH$_3$PbI$_3$ (X= I, Br). \emph{Molecules}
  \textbf{2019}, \emph{24}, 626\relax
\mciteBstWouldAddEndPuncttrue
\mciteSetBstMidEndSepPunct{\mcitedefaultmidpunct}
{\mcitedefaultendpunct}{\mcitedefaultseppunct}\relax
\EndOfBibitem
\bibitem[P\'{e}rez-Osorio \latin{et~al.}(2018)P\'{e}rez-Osorio, Lin, Phillips,
  Milot, Herz, Johnston, and Giustino]{perez2018raman}
P\'{e}rez-Osorio,~M.~A.; Lin,~Q.; Phillips,~R.~T.; Milot,~R.~L.; Herz,~L.~M.;
  Johnston,~M.~B.; Giustino,~F. Raman spectrum of the organic--inorganic halide
  perovskite CH$_3$NH$_3$PbI$_3$ from first principles and high-resolution
  low-temperature Raman measurements. \emph{J. Phys. Chem. C} \textbf{2018},
  \emph{122}, 21703--21717\relax
\mciteBstWouldAddEndPuncttrue
\mciteSetBstMidEndSepPunct{\mcitedefaultmidpunct}
{\mcitedefaultendpunct}{\mcitedefaultseppunct}\relax
\EndOfBibitem
\bibitem[P{\'e}rez-Osorio \latin{et~al.}(2015)P{\'e}rez-Osorio, Milot, Filip,
  Patel, Herz, Johnston, and Giustino]{perez2015vibrational}
P{\'e}rez-Osorio,~M.~A.; Milot,~R.~L.; Filip,~M.~R.; Patel,~J.~B.; Herz,~L.~M.;
  Johnston,~M.~B.; Giustino,~F. Vibrational properties of the
  organic--inorganic halide perovskite CH$_3$NH$_3$PbI$_3$ from theory and
  experiment: factor group analysis, first-principles calculations, and
  low-temperature infrared spectra. \emph{J. Phys. Chem. C} \textbf{2015},
  \emph{119}, 25703--25718\relax
\mciteBstWouldAddEndPuncttrue
\mciteSetBstMidEndSepPunct{\mcitedefaultmidpunct}
{\mcitedefaultendpunct}{\mcitedefaultseppunct}\relax
\EndOfBibitem
\bibitem[Glaser \latin{et~al.}(2015)Glaser, M\"{u}ller, Sendner, Krekeler,
  Semonin, Hull, Yaffe, Owen, Kowalsky, Pucci, \latin{et~al.}
  others]{glaser2015infrared}
Glaser,~T.; M\"{u}ller,~C.; Sendner,~M.; Krekeler,~C.; Semonin,~O.~E.;
  Hull,~T.~D.; Yaffe,~O.; Owen,~J.~S.; Kowalsky,~W.; Pucci,~A. \latin{et~al.}
  Infrared spectroscopic study of vibrational modes in methylammonium lead
  halide perovskites. \emph{J. Phys. Chem. Lett.} \textbf{2015}, \emph{6},
  2913--2918\relax
\mciteBstWouldAddEndPuncttrue
\mciteSetBstMidEndSepPunct{\mcitedefaultmidpunct}
{\mcitedefaultendpunct}{\mcitedefaultseppunct}\relax
\EndOfBibitem
\bibitem[Leguy \latin{et~al.}(2016)Leguy, Go{\~n}i, Frost, Skelton, Brivio,
  Rodr{\'\i}guez-Mart{\'\i}nez, Weber, Pallipurath, Alonso, Campoy-Quiles,
  \latin{et~al.} others]{leguy2016dynamic}
Leguy,~A.~M.; Go{\~n}i,~A.~R.; Frost,~J.~M.; Skelton,~J.; Brivio,~F.;
  Rodr{\'\i}guez-Mart{\'\i}nez,~X.; Weber,~O.~J.; Pallipurath,~A.;
  Alonso,~M.~I.; Campoy-Quiles,~M. \latin{et~al.}  Dynamic disorder, phonon
  lifetimes, and the assignment of modes to the vibrational spectra of
  methylammonium lead halide perovskites. \emph{Phys. Chem. Chem. Phys.}
  \textbf{2016}, \emph{18}, 27051--27066\relax
\mciteBstWouldAddEndPuncttrue
\mciteSetBstMidEndSepPunct{\mcitedefaultmidpunct}
{\mcitedefaultendpunct}{\mcitedefaultseppunct}\relax
\EndOfBibitem
\bibitem[Chen \latin{et~al.}(2016)Chen, Liu, Kim, Liu, Yang, Yue, Ren, Zhu,
  Liu, Park, and Zhang]{Y.Zhang}
Chen,~Q.; Liu,~H.; Kim,~H.-S.; Liu,~Y.; Yang,~M.; Yue,~N.; Ren,~G.; Zhu,~K.;
  Liu,~S.; Park,~N.-G. \latin{et~al.}  Multiple-stage structure transformation
  of organic-inorganic hybrid perovskite
  ${\mathrm{CH}}_{3}{\mathrm{NH}}_{3}{\mathrm{PbI}}_{3}$. \emph{Phys. Rev. X}
  \textbf{2016}, \emph{6}, 031042\relax
\mciteBstWouldAddEndPuncttrue
\mciteSetBstMidEndSepPunct{\mcitedefaultmidpunct}
{\mcitedefaultendpunct}{\mcitedefaultseppunct}\relax
\EndOfBibitem
\bibitem[Matsuishi \latin{et~al.}(2004)Matsuishi, Ishihara, Onari, Chang, and
  Park]{matsuishi2004optical}
Matsuishi,~K.; Ishihara,~T.; Onari,~S.; Chang,~Y.; Park,~C. Optical properties
  and structural phase transitions of lead-halide based inorganic--organic 3D
  and 2D perovskite semiconductors under high pressure. \emph{phys. stat.
  sol. (b)} \textbf{2004}, \emph{241}, 3328--3333\relax
\mciteBstWouldAddEndPuncttrue
\mciteSetBstMidEndSepPunct{\mcitedefaultmidpunct}
{\mcitedefaultendpunct}{\mcitedefaultseppunct}\relax
\EndOfBibitem
\bibitem[Ge \latin{et~al.}(2018)Ge, Hu, Wu, Tan, Chen, Wang, Shi, and
  Feng]{Chunyu}
Ge,~C.; Hu,~M.; Wu,~P.; Tan,~Q.; Chen,~Z.; Wang,~Y.; Shi,~J.; Feng,~J. Ultralow
  thermal conductivity and ultrahigh thermal expansion of single-crystal
  organic--inorganic hybrid perovskite CH$_3$NH$_3$PbX$_3$ (X= Cl, Br, I).
  \emph{J. Phys. Chem. C} \textbf{2018}, \emph{122}, 15973--15978\relax
\mciteBstWouldAddEndPuncttrue
\mciteSetBstMidEndSepPunct{\mcitedefaultmidpunct}
{\mcitedefaultendpunct}{\mcitedefaultseppunct}\relax
\EndOfBibitem
\bibitem[Heiderhoff \latin{et~al.}(2017)Heiderhoff, Haeger, Pourdavoud, Hu,
  Al-Khafaji, Mayer, Chen, Scheer, and Riedl]{heiderhoff2017thermal}
Heiderhoff,~R.; Haeger,~T.; Pourdavoud,~N.; Hu,~T.; Al-Khafaji,~M.; Mayer,~A.;
  Chen,~Y.; Scheer,~H.-C.; Riedl,~T. Thermal conductivity of methylammonium
  lead halide perovskite single crystals and thin films: A comparative study.
  \emph{J. Phys. Chem. C} \textbf{2017}, \emph{121}, 28306--28311\relax
\mciteBstWouldAddEndPuncttrue
\mciteSetBstMidEndSepPunct{\mcitedefaultmidpunct}
{\mcitedefaultendpunct}{\mcitedefaultseppunct}\relax
\EndOfBibitem
\bibitem[Anastassakis \latin{et~al.}(1993)Anastassakis, Pinczuk, Burstein,
  Pollak, and Cardona]{anastassakis1993effect}
Anastassakis,~E.; Pinczuk,~A.; Burstein,~E.; Pollak,~F.; Cardona,~M. Effect of
  static uniaxial stress on the Raman spectrum of silicon. \emph{Solid State
  Commun.} \textbf{1993}, \emph{88}, 1053--1058\relax
\mciteBstWouldAddEndPuncttrue
\mciteSetBstMidEndSepPunct{\mcitedefaultmidpunct}
{\mcitedefaultendpunct}{\mcitedefaultseppunct}\relax
\EndOfBibitem
\bibitem[Fabini \latin{et~al.}(2020)Fabini, Seshadri, and
  Kanatzidis]{fabini2020underappreciated}
Fabini,~D.~H.; Seshadri,~R.; Kanatzidis,~M.~G. The underappreciated lone pair
  in halide perovskites underpins their unusual properties. \emph{MRS Bull.}
  \textbf{2020}, \emph{45}, 467--477\relax
\mciteBstWouldAddEndPuncttrue
\mciteSetBstMidEndSepPunct{\mcitedefaultmidpunct}
{\mcitedefaultendpunct}{\mcitedefaultseppunct}\relax
\EndOfBibitem
\bibitem[Zhu and Ertekin(2019)Zhu, and Ertekin]{zhu2019mixed}
Zhu,~T.; Ertekin,~E. Mixed phononic and non-phononic transport in hybrid lead
  halide perovskites: glass-crystal duality, dynamical disorder, and
  anharmonicity. \emph{Energy Environ. Sci.} \textbf{2019},
  \emph{12}, 216--229\relax
\mciteBstWouldAddEndPuncttrue
\mciteSetBstMidEndSepPunct{\mcitedefaultmidpunct}
{\mcitedefaultendpunct}{\mcitedefaultseppunct}\relax
\EndOfBibitem
\bibitem[Rolston \latin{et~al.}(2018)Rolston, Bush, Printz, Gold-Parker, Ding,
  Toney, McGehee, and Dauskardt]{rolston2018engineering}
Rolston,~N.; Bush,~K.~A.; Printz,~A.~D.; Gold-Parker,~A.; Ding,~Y.;
  Toney,~M.~F.; McGehee,~M.~D.; Dauskardt,~R.~H. Engineering stress in
  perovskite solar cells to improve stability. \emph{Adv. Energy Mater.}
  \textbf{2018}, \emph{8}, 1802139\relax
\mciteBstWouldAddEndPuncttrue
\mciteSetBstMidEndSepPunct{\mcitedefaultmidpunct}
{\mcitedefaultendpunct}{\mcitedefaultseppunct}\relax
\EndOfBibitem
\bibitem[Pistor \latin{et~al.}(2016)Pistor, Ruiz, Cabot, and
  Izquierdo-Roca]{Pistor}
Pistor,~P.; Ruiz,~A.; Cabot,~A.; Izquierdo-Roca,~V. Advanced Raman spectroscopy
  of methylammonium lead iodide: Development of a non-destructive
  characterisation methodology. \emph{Sci. Rep.} \textbf{2016}, \emph{6},
  35973\relax
\mciteBstWouldAddEndPuncttrue
\mciteSetBstMidEndSepPunct{\mcitedefaultmidpunct}
{\mcitedefaultendpunct}{\mcitedefaultseppunct}\relax
\EndOfBibitem
\bibitem[De~Wolf(1996)]{de1996micro}
De~Wolf,~I. Micro-Raman spectroscopy to study local mechanical stress in
  silicon integrated circuits. \emph{Semicond. Sci. Technol.} \textbf{1996},
  \emph{11}, 139\relax
\mciteBstWouldAddEndPuncttrue
\mciteSetBstMidEndSepPunct{\mcitedefaultmidpunct}
{\mcitedefaultendpunct}{\mcitedefaultseppunct}\relax
\EndOfBibitem
\bibitem[Rao \latin{et~al.}(2019)Rao, Islam, Singh, Berry, Kawakami, Maruyama,
  and Katoch]{rao2019spectroscopic}
Rao,~R.; Islam,~A.~E.; Singh,~S.; Berry,~R.; Kawakami,~R.~K.; Maruyama,~B.;
  Katoch,~J. Spectroscopic evaluation of charge-transfer doping and strain in
  graphene/MoS$_2$ heterostructures. \emph{Phys. Rev. B} \textbf{2019},
  \emph{99}, 195401\relax
\mciteBstWouldAddEndPuncttrue
\mciteSetBstMidEndSepPunct{\mcitedefaultmidpunct}
{\mcitedefaultendpunct}{\mcitedefaultseppunct}\relax
\EndOfBibitem
\bibitem[Strubbe \latin{et~al.}(2015)Strubbe, Johlin, Kirkpatrick, Buonassisi,
  and Grossman]{Strubbe}
Strubbe,~D.~A.; Johlin,~E.~C.; Kirkpatrick,~T.~R.; Buonassisi,~T.;
  Grossman,~J.~C. Stress effects on the Raman spectrum of an amorphous
  material: Theory and experiment on $a$-Si:H. \emph{Phys. Rev. B}
  \textbf{2015}, \emph{92}, 241202(R)\relax
\mciteBstWouldAddEndPuncttrue
\mciteSetBstMidEndSepPunct{\mcitedefaultmidpunct}
{\mcitedefaultendpunct}{\mcitedefaultseppunct}\relax
\EndOfBibitem
\bibitem[Su \latin{et~al.}(2016)Su, Kumar, Dai, and Roy]{su2016nanoscale}
Su,~W.; Kumar,~N.; Dai,~N.; Roy,~D. Nanoscale mapping of intrinsic defects in
  single-layer graphene using tip-enhanced Raman spectroscopy. \emph{Chem.
  Commun.} \textbf{2016}, \emph{52}, 8227--8230\relax
\mciteBstWouldAddEndPuncttrue
\mciteSetBstMidEndSepPunct{\mcitedefaultmidpunct}
{\mcitedefaultendpunct}{\mcitedefaultseppunct}\relax
\EndOfBibitem
\bibitem[Lyu \latin{et~al.}(2019)Lyu, Li, Jiang, Shan, Hu, Deng, Ying, Wang,
  Zhang, Bechtel, \latin{et~al.} others]{lyu2019phonon}
Lyu,~B.; Li,~H.; Jiang,~L.; Shan,~W.; Hu,~C.; Deng,~A.; Ying,~Z.; Wang,~L.;
  Zhang,~Y.; Bechtel,~H.~A. \latin{et~al.}  Phonon polariton-assisted infrared
  nanoimaging of local strain in hexagonal boron nitride. \emph{Nano Lett.}
  \textbf{2019}, \emph{19}, 1982--1989\relax
\mciteBstWouldAddEndPuncttrue
\mciteSetBstMidEndSepPunct{\mcitedefaultmidpunct}
{\mcitedefaultendpunct}{\mcitedefaultseppunct}\relax
\EndOfBibitem
\bibitem[Smith \latin{et~al.}(2019)Smith, Nowadnick, Fan, Khatib, Lim, Gao,
  Harms, Neal, Kirkland, Martin, \latin{et~al.} others]{smith2019infrared}
Smith,~K.; Nowadnick,~E.; Fan,~S.; Khatib,~O.; Lim,~S.~J.; Gao,~B.; Harms,~N.;
  Neal,~S.; Kirkland,~J.; Martin,~M. \latin{et~al.}  Infrared nano-spectroscopy
  of ferroelastic domain walls in hybrid improper ferroelectric
  Ca$_3$Ti$_2$O$_7$. \emph{Nat. Commun.} \textbf{2019}, \emph{10}, 1--9\relax
\mciteBstWouldAddEndPuncttrue
\mciteSetBstMidEndSepPunct{\mcitedefaultmidpunct}
{\mcitedefaultendpunct}{\mcitedefaultseppunct}\relax
\EndOfBibitem
\bibitem[Baroni \latin{et~al.}(2001)Baroni, de~Gironcoli, Dal~Corso, and
  Giannozzi]{Baroni.RevModPhys.73.515}
Baroni,~S.; de~Gironcoli,~S.; Dal~Corso,~A.; Giannozzi,~P. Phonons and related
  crystal properties from density-functional perturbation theory. \emph{Rev.
  Mod. Phys.} \textbf{2001}, \emph{73}, 515--562\relax
\mciteBstWouldAddEndPuncttrue
\mciteSetBstMidEndSepPunct{\mcitedefaultmidpunct}
{\mcitedefaultendpunct}{\mcitedefaultseppunct}\relax
\EndOfBibitem
\bibitem[Giannozzi \latin{et~al.}(2017)Giannozzi, Andreussi, Brumme, Bunau,
  Nardelli, Calandra, Car, Cavazzoni, Ceresoli, Cococcioni, \latin{et~al.}
  others]{giannozzi2017advanced}
Giannozzi,~P.; Andreussi,~O.; Brumme,~T.; Bunau,~O.; Nardelli,~M.~B.;
  Calandra,~M.; Car,~R.; Cavazzoni,~C.; Ceresoli,~D.; Cococcioni,~M.
  \latin{et~al.}  Advanced capabilities for materials modelling with Quantum
  ESPRESSO. \emph{J. Phys.: Condens. Matt.} \textbf{2017}, \emph{29},
  465901\relax
\mciteBstWouldAddEndPuncttrue
\mciteSetBstMidEndSepPunct{\mcitedefaultmidpunct}
{\mcitedefaultendpunct}{\mcitedefaultseppunct}\relax
\EndOfBibitem
\bibitem[Giannozzi \latin{et~al.}(2009)Giannozzi, Baroni, Bonini, Calandra,
  Car, Cavazzoni, Ceresoli, Chiarotti, Cococcioni, Dabo, Corso, de~Gironcoli,
  Fabris, Fratesi, Gebauer, Gerstmann, Gougoussis, Kokalj, Lazzeri,
  Martin-Samos, Marzari, Mauri, Mazzarello, Paolini, Pasquarello, Paulatto,
  Sbraccia, Scandolo, Sclauzero, Seitsonen, Smogunov, Umari, and
  Wentzcovitch]{Giannozzi}
Giannozzi,~P.; Baroni,~S.; Bonini,~N.; Calandra,~M.; Car,~R.; Cavazzoni,~C.;
  Ceresoli,~D.; Chiarotti,~G.~L.; Cococcioni,~M.; Dabo,~I. \latin{et~al.}
  {Quantum} {ESPRESSO}: a modular and open-source software project for quantum
  simulations of materials. \emph{J. Phys.: Condens. Matt.} \textbf{2009},
  \emph{21}, 395502\relax
\mciteBstWouldAddEndPuncttrue
\mciteSetBstMidEndSepPunct{\mcitedefaultmidpunct}
{\mcitedefaultendpunct}{\mcitedefaultseppunct}\relax
\EndOfBibitem
\bibitem[Baroni \latin{et~al.}(2005)Baroni, Dal~Corso, De~Gironcoli, Giannozzi,
  Cavazzoni, Ballabio, Scandolo, Chiarotti, Focher, Pasquarello, \latin{et~al.}
  others]{QE}
Baroni,~S.; Dal~Corso,~A.; De~Gironcoli,~S.; Giannozzi,~P.; Cavazzoni,~C.;
  Ballabio,~G.; Scandolo,~S.; Chiarotti,~G.; Focher,~P.; Pasquarello,~A.
  \latin{et~al.}  Quantum ESPRESSO: open-source package for research in
  electronic structure, simulation, and optimization. \emph{Code available from
  http://www.quantum-espresso.org} \textbf{2005} \relax
\mciteBstWouldAddEndPunctfalse
\mciteSetBstMidEndSepPunct{\mcitedefaultmidpunct}
{}{\mcitedefaultseppunct}\relax
\EndOfBibitem
\bibitem[Perdew and Zunger(1981)Perdew, and Zunger]{LDA_PhysRevB.23.5048}
Perdew,~J.~P.; Zunger,~A. Self-interaction correction to density-functional
  approximations for many-electron systems. \emph{Phys. Rev. B} \textbf{1981},
  \emph{23}, 5048--5079\relax
\mciteBstWouldAddEndPuncttrue
\mciteSetBstMidEndSepPunct{\mcitedefaultmidpunct}
{\mcitedefaultendpunct}{\mcitedefaultseppunct}\relax
\EndOfBibitem
\bibitem[Perdew and Wang(1992)Perdew, and Wang]{perdew1992accurate}
Perdew,~J.~P.; Wang,~Y. Accurate and simple analytic representation of the
  electron-gas correlation energy. \emph{Phys. Rev. B} \textbf{1992},
  \emph{45}, 13244\relax
\mciteBstWouldAddEndPuncttrue
\mciteSetBstMidEndSepPunct{\mcitedefaultmidpunct}
{\mcitedefaultendpunct}{\mcitedefaultseppunct}\relax
\EndOfBibitem
\bibitem[Hamann(2013)]{Hamann}
Hamann,~D.~R. Optimized norm-conserving Vanderbilt pseudopotentials.
  \emph{Phys. Rev. B} \textbf{2013}, \emph{88}, 085117\relax
\mciteBstWouldAddEndPuncttrue
\mciteSetBstMidEndSepPunct{\mcitedefaultmidpunct}
{\mcitedefaultendpunct}{\mcitedefaultseppunct}\relax
\EndOfBibitem
\bibitem[Van~Setten \latin{et~al.}(2018)Van~Setten, Giantomassi, Bousquet,
  Verstraete, Hamann, Gonze, and Rignanese]{van2018pseudodojo}
Van~Setten,~M.; Giantomassi,~M.; Bousquet,~E.; Verstraete,~M.~J.;
  Hamann,~D.~R.; Gonze,~X.; Rignanese,~G.-M. The PseudoDojo: Training and
  grading a 85 element optimized norm-conserving pseudopotential table.
  \emph{Comput. Phys. Commun.} \textbf{2018}, \emph{226}, 39--54\relax
\mciteBstWouldAddEndPuncttrue
\mciteSetBstMidEndSepPunct{\mcitedefaultmidpunct}
{\mcitedefaultendpunct}{\mcitedefaultseppunct}\relax
\EndOfBibitem
\bibitem[pse()]{pseudodojo}
The PseudoDojo: Training and grading a 85 element optimized norm-conserving
  pseudopotential table. \url{http://www.pseudo-dojo.org} (accessed February
  09, 2020)\relax
\mciteBstWouldAddEndPuncttrue
\mciteSetBstMidEndSepPunct{\mcitedefaultmidpunct}
{\mcitedefaultendpunct}{\mcitedefaultseppunct}\relax
\EndOfBibitem
\bibitem[Str()]{Structure}
Initial structure of MAPI.
  \url{https://github.com/WMD-group/hybrid-perovskites/blob/master/2015_ch3nh3pbi3_phonons_PBEsol/CH3NH3PbI3_cubic.cif}
  (accessed February 09, 2020)\relax
\mciteBstWouldAddEndPuncttrue
\mciteSetBstMidEndSepPunct{\mcitedefaultmidpunct}
{\mcitedefaultendpunct}{\mcitedefaultseppunct}\relax
\EndOfBibitem
\bibitem[Weller \latin{et~al.}(2015)Weller, Weber, Henry, {Di Pumpo}, and
  Hansen]{PND}
Weller,~M.~T.; Weber,~O.~J.; Henry,~P.~F.; {Di Pumpo},~A.~M.; Hansen,~T.~C.
  {Complete structure and cation orientation in the perovskite photovoltaic
  methylammonium lead iodide between 100 and 352 K}. \emph{Chem. Commun.}
  \textbf{2015}, \emph{51}, 4180--4183\relax
\mciteBstWouldAddEndPuncttrue
\mciteSetBstMidEndSepPunct{\mcitedefaultmidpunct}
{\mcitedefaultendpunct}{\mcitedefaultseppunct}\relax
\EndOfBibitem
\bibitem[Baikie \latin{et~al.}(2013)Baikie, Fang, Kadro, Schreyer, Wei,
  Mhaisalkar, Graetzel, and White]{baikie2013synthesis}
Baikie,~T.; Fang,~Y.; Kadro,~J.~M.; Schreyer,~M.; Wei,~F.; Mhaisalkar,~S.~G.;
  Graetzel,~M.; White,~T.~J. Synthesis and crystal chemistry of the hybrid
  perovskite CH$_3$NH$_3$PbI$_3$ for solid-state sensitised solar cell
  applications. \emph{J. Mater. Chem. A} \textbf{2013}, \emph{1},
  5628--5641\relax
\mciteBstWouldAddEndPuncttrue
\mciteSetBstMidEndSepPunct{\mcitedefaultmidpunct}
{\mcitedefaultendpunct}{\mcitedefaultseppunct}\relax
\EndOfBibitem
\bibitem[Stoumpos \latin{et~al.}(2013)Stoumpos, Malliakas, and
  Kanatzidis]{stoumpos2013semiconducting}
Stoumpos,~C.~C.; Malliakas,~C.~D.; Kanatzidis,~M.~G. Semiconducting tin and
  lead iodide perovskites with organic cations: phase transitions, high
  mobilities, and near-infrared photoluminescent properties. \emph{Inorg.
  Chem.} \textbf{2013}, \emph{52}, 9019--9038\relax
\mciteBstWouldAddEndPuncttrue
\mciteSetBstMidEndSepPunct{\mcitedefaultmidpunct}
{\mcitedefaultendpunct}{\mcitedefaultseppunct}\relax
\EndOfBibitem
\bibitem[Leguy \latin{et~al.}(2015)Leguy, Frost, McMahon, Sakai, Kockelmann,
  Law, Li, Foglia, Walsh, O'Regan, \latin{et~al.} others]{leguy2015dynamics}
Leguy,~A.~M.; Frost,~J.~M.; McMahon,~A.~P.; Sakai,~V.~G.; Kockelmann,~W.;
  Law,~C.; Li,~X.; Foglia,~F.; Walsh,~A.; O'Regan,~B.~C. \latin{et~al.}  The
  dynamics of methylammonium ions in hybrid organic--inorganic perovskite solar
  cells. \emph{Nat. Commun.} \textbf{2015}, \emph{6}, 7124\relax
\mciteBstWouldAddEndPuncttrue
\mciteSetBstMidEndSepPunct{\mcitedefaultmidpunct}
{\mcitedefaultendpunct}{\mcitedefaultseppunct}\relax
\EndOfBibitem
\bibitem[Whalley \latin{et~al.}(2017)Whalley, Frost, Jung, and
  Walsh]{Whalley2017}
Whalley,~L.~D.; Frost,~J.~M.; Jung,~Y.-K.; Walsh,~A. Perspective: Theory and
  simulation of hybrid halide perovskites. \emph{J. Chem. Phys.} \textbf{2017},
  \emph{146}, 220901\relax
\mciteBstWouldAddEndPuncttrue
\mciteSetBstMidEndSepPunct{\mcitedefaultmidpunct}
{\mcitedefaultendpunct}{\mcitedefaultseppunct}\relax
\EndOfBibitem
\bibitem[McKechnie \latin{et~al.}(2018)McKechnie, Frost, Pashov, Azarhoosh,
  Walsh, and {Van Schilfgaarde}]{McKechnie2018}
McKechnie,~S.; Frost,~J.~M.; Pashov,~D.; Azarhoosh,~P.; Walsh,~A.; {Van
  Schilfgaarde},~M. {Dynamic symmetry breaking and spin splitting in metal
  halide perovskites}. \emph{Phys. Rev. B} \textbf{2018}, \emph{98},
  085108\relax
\mciteBstWouldAddEndPuncttrue
\mciteSetBstMidEndSepPunct{\mcitedefaultmidpunct}
{\mcitedefaultendpunct}{\mcitedefaultseppunct}\relax
\EndOfBibitem
\bibitem[Giorgi \latin{et~al.}(2013)Giorgi, Fujisawa, Segawa, and
  Yamashita]{Giorgi2013}
Giorgi,~G.; Fujisawa,~J.-i.~I.; Segawa,~H.; Yamashita,~K. {Small photocarrier
  effective masses featuring ambipolar transport in methylammonium lead iodide
  perovskite: A density functional analysis}. \emph{J. Phys. Chem. Lett.}
  \textbf{2013}, \emph{4}, 4213--4216\relax
\mciteBstWouldAddEndPuncttrue
\mciteSetBstMidEndSepPunct{\mcitedefaultmidpunct}
{\mcitedefaultendpunct}{\mcitedefaultseppunct}\relax
\EndOfBibitem
\bibitem[Perdew \latin{et~al.}(1996)Perdew, Burke, and
  Ernzerhof]{PBE_PhysRevLett.77.3865}
Perdew,~J.~P.; Burke,~K.; Ernzerhof,~M. Generalized gradient approximation made
  simple. \emph{Phys. Rev. Lett.} \textbf{1996}, \emph{77}, 3865--3868\relax
\mciteBstWouldAddEndPuncttrue
\mciteSetBstMidEndSepPunct{\mcitedefaultmidpunct}
{\mcitedefaultendpunct}{\mcitedefaultseppunct}\relax
\EndOfBibitem
\bibitem[Perdew \latin{et~al.}(2008)Perdew, Ruzsinszky, Csonka, Vydrov,
  Scuseria, Constantin, Zhou, and Burke]{PBESol_PhysRevLett.100.136406}
Perdew,~J.~P.; Ruzsinszky,~A.; Csonka,~G.~I.; Vydrov,~O.~A.; Scuseria,~G.~E.;
  Constantin,~L.~A.; Zhou,~X.; Burke,~K. Restoring the density-gradient
  expansion for exchange in solids and surfaces. \emph{Phys. Rev. Lett.}
  \textbf{2008}, \emph{100}, 136406\relax
\mciteBstWouldAddEndPuncttrue
\mciteSetBstMidEndSepPunct{\mcitedefaultmidpunct}
{\mcitedefaultendpunct}{\mcitedefaultseppunct}\relax
\EndOfBibitem
\bibitem[Parker~Jr \latin{et~al.}(1967)Parker~Jr, Feldman, and
  Ashkin]{parker1967raman}
Parker~Jr,~J.; Feldman,~D.; Ashkin,~M. Raman scattering by silicon and
  germanium. \emph{Phys. Rev.} \textbf{1967}, \emph{155}, 712\relax
\mciteBstWouldAddEndPuncttrue
\mciteSetBstMidEndSepPunct{\mcitedefaultmidpunct}
{\mcitedefaultendpunct}{\mcitedefaultseppunct}\relax
\EndOfBibitem
\bibitem[Anastassakis \latin{et~al.}(1990)Anastassakis, Cantarero, and
  Cardona]{anastassakis1990piezo}
Anastassakis,~E.; Cantarero,~A.; Cardona,~M. Piezo-Raman measurements and
  anharmonic parameters in silicon and diamond. \emph{Phys. Rev. B}
  \textbf{1990}, \emph{41}, 7529\relax
\mciteBstWouldAddEndPuncttrue
\mciteSetBstMidEndSepPunct{\mcitedefaultmidpunct}
{\mcitedefaultendpunct}{\mcitedefaultseppunct}\relax
\EndOfBibitem
\bibitem[Giustino(2014)]{giustino2014materials}
Giustino,~F. \emph{Materials modelling using density functional theory:
  properties and predictions}; Oxford University Press, 2014\relax
\mciteBstWouldAddEndPuncttrue
\mciteSetBstMidEndSepPunct{\mcitedefaultmidpunct}
{\mcitedefaultendpunct}{\mcitedefaultseppunct}\relax
\EndOfBibitem
\bibitem[Feng(2014)]{Feng2014}
Feng,~J. Mechanical properties of hybrid organic-inorganic CH$_3$NH$_3$BX$_3$
  (B=Sn, Pb; X=Br, I) perovskites for solar cell absorbers. \emph{APL Mater.}
  \textbf{2014}, \emph{2}, 081801\relax
\mciteBstWouldAddEndPuncttrue
\mciteSetBstMidEndSepPunct{\mcitedefaultmidpunct}
{\mcitedefaultendpunct}{\mcitedefaultseppunct}\relax
\EndOfBibitem
\bibitem[Lazzeri and Mauri(2003)Lazzeri, and Mauri]{lazzeri2003first}
Lazzeri,~M.; Mauri,~F. First-principles calculation of vibrational Raman
  spectra in large systems: Signature of small rings in crystalline SiO$_2$.
  \emph{Phys. Rev. Lett.} \textbf{2003}, \emph{90}, 036401\relax
\mciteBstWouldAddEndPuncttrue
\mciteSetBstMidEndSepPunct{\mcitedefaultmidpunct}
{\mcitedefaultendpunct}{\mcitedefaultseppunct}\relax
\EndOfBibitem
\bibitem[Mounet(2005)]{Mounet}
Mounet,~N. Structural, vibrational and thermodynamic properties of carbon
  allotropes from first-principles: diamond, graphite, and nanotubes.
  \emph{Masters Thesis, Massachusetts Institute of Technology} \textbf{2005} \relax
\mciteBstWouldAddEndPunctfalse
\mciteSetBstMidEndSepPunct{\mcitedefaultmidpunct}
{}{\mcitedefaultseppunct}\relax
\EndOfBibitem
\bibitem[Vo{\v{c}}adlo and Price(1994)Vo{\v{c}}adlo, and Price]{Vocadlo1994}
Vo{\v{c}}adlo,~N.~L.; Price,~G.~D. {The Gr{\"{u}}neisen parameter -- computer
  calculations via lattice dynamics}. \emph{Phys. Earth Planet. Inter.}
  \textbf{1994}, \emph{82}, 261--270\relax
\mciteBstWouldAddEndPuncttrue
\mciteSetBstMidEndSepPunct{\mcitedefaultmidpunct}
{\mcitedefaultendpunct}{\mcitedefaultseppunct}\relax
\EndOfBibitem
\bibitem[Fennie and Rabe(2003)Fennie, and Rabe]{Rabe}
Fennie,~C.~J.; Rabe,~K.~M. Structural and dielectric properties of
  ${\mathrm{Sr}}_{2}{\mathrm{TiO}}_{4}$ from first principles. \emph{Phys. Rev.
  B} \textbf{2003}, \emph{68}, 184111\relax
\mciteBstWouldAddEndPuncttrue
\mciteSetBstMidEndSepPunct{\mcitedefaultmidpunct}
{\mcitedefaultendpunct}{\mcitedefaultseppunct}\relax
\EndOfBibitem
\bibitem[Ganesan \latin{et~al.}(1970)Ganesan, Maradudin, and
  Oitmaa]{ganesan1970lattice}
Ganesan,~S.; Maradudin,~A.; Oitmaa,~J. A lattice theory of morphic effects in
  crystals of the diamond structure. \emph{Ann. Phys.} \textbf{1970},
  \emph{56}, 556--594\relax
\mciteBstWouldAddEndPuncttrue
\mciteSetBstMidEndSepPunct{\mcitedefaultmidpunct}
{\mcitedefaultendpunct}{\mcitedefaultseppunct}\relax
\EndOfBibitem
\bibitem[Xue \latin{et~al.}(2020)Xue, Hou, Liu, Wei, Chen, Huang, Li, Sun,
  Proppe, Dong, \latin{et~al.} others]{xue2020regulating}
Xue,~D.-J.; Hou,~Y.; Liu,~S.-C.; Wei,~M.; Chen,~B.; Huang,~Z.; Li,~Z.; Sun,~B.;
  Proppe,~A.~H.; Dong,~Y. \latin{et~al.}  Regulating strain in perovskite thin
  films through charge-transport layers. \emph{Nat. Commun.} \textbf{2020},
  \emph{11}, 1--8\relax
\mciteBstWouldAddEndPuncttrue
\mciteSetBstMidEndSepPunct{\mcitedefaultmidpunct}
{\mcitedefaultendpunct}{\mcitedefaultseppunct}\relax
\EndOfBibitem
\bibitem[Rappich \latin{et~al.}(2020)Rappich, Lang, Brus, Shargaieva, Dittrich,
  and Nickel]{rappich2020light}
Rappich,~J.; Lang,~F.; Brus,~V.~V.; Shargaieva,~O.; Dittrich,~T.; Nickel,~N.~H.
  Light-induced defect generation in CH$_3$NH$_3$PbI$_3$ thin films and single
  crystals. \emph{Solar RRL} \textbf{2020}, \emph{4}, 1900216\relax
\mciteBstWouldAddEndPuncttrue
\mciteSetBstMidEndSepPunct{\mcitedefaultmidpunct}
{\mcitedefaultendpunct}{\mcitedefaultseppunct}\relax
\EndOfBibitem
\bibitem[Kang and Biswas(2017)Kang, and Biswas]{kang2017preferential}
Kang,~B.; Biswas,~K. Preferential CH$_3$NH$_3^{+}$ alignment and octahedral
  tilting affect charge localization in cubic phase CH$_3$NH$_3$PbI$_3$.
  \emph{J. Phys. Chem. C} \textbf{2017}, \emph{121}, 8319--8326\relax
\mciteBstWouldAddEndPuncttrue
\mciteSetBstMidEndSepPunct{\mcitedefaultmidpunct}
{\mcitedefaultendpunct}{\mcitedefaultseppunct}\relax
\EndOfBibitem
\bibitem[Wu \latin{et~al.}(2018)Wu, Wang, Ercius, Wright, Leppert-Simenauer,
  Burke, Dubey, Dogare, and Pettes]{wu2018giant}
Wu,~W.; Wang,~J.; Ercius,~P.; Wright,~N.~C.; Leppert-Simenauer,~D.~M.;
  Burke,~R.~A.; Dubey,~M.; Dogare,~A.~M.; Pettes,~M.~T. Giant
  mechano-optoelectronic effect in an atomically thin semiconductor. \emph{Nano
  Lett.} \textbf{2018}, \emph{18}, 2351--2357\relax
\mciteBstWouldAddEndPuncttrue
\mciteSetBstMidEndSepPunct{\mcitedefaultmidpunct}
{\mcitedefaultendpunct}{\mcitedefaultseppunct}\relax
\EndOfBibitem
\bibitem[Bonini \latin{et~al.}(2007)Bonini, Lazzeri, Marzari, and
  Mauri]{bonini2007phonon}
Bonini,~N.; Lazzeri,~M.; Marzari,~N.; Mauri,~F. Phonon anharmonicities in
  graphite and graphene. \emph{Phys. Rev. Lett.} \textbf{2007}, \emph{99},
  176802\relax
\mciteBstWouldAddEndPuncttrue
\mciteSetBstMidEndSepPunct{\mcitedefaultmidpunct}
{\mcitedefaultendpunct}{\mcitedefaultseppunct}\relax
\EndOfBibitem
\bibitem[Deringer \latin{et~al.}(2014)Deringer, Stoffel, and
  Dronskowski]{Deringer}
Deringer,~V.~L.; Stoffel,~R.~P.; Dronskowski,~R. Vibrational and thermodynamic
  properties of GeSe in the quasiharmonic approximation. \emph{Phys. Rev. B}
  \textbf{2014}, \emph{89}, 094303\relax
\mciteBstWouldAddEndPuncttrue
\mciteSetBstMidEndSepPunct{\mcitedefaultmidpunct}
{\mcitedefaultendpunct}{\mcitedefaultseppunct}\relax
\EndOfBibitem
\end{mcitethebibliography}
\end{document}